\documentclass[onecolumn,nofootinbib,showkeys]{revtex4}
\usepackage{amsmath,amssymb}
\usepackage{graphicx}
\usepackage{color}

\begin{document}

\title{Photon sphere and shadow of a time-dependent black hole \\ described by a Vaidya metric}

\author{Jay Solanki}
\email{jay565109@gmail.com}
\affiliation{Sardar Vallabhbhai National Institute of Technology,\\ Surat - 395007, Gujarat, India}

\author{Volker Perlick}
\email{perlick@zarm.uni-bremen.de}
\affiliation{ZARM, University of Bremen, 28359 Bremen, Germany}

\begin{abstract}
    In this paper we derive exact analytical formulas for the evolution of the photon sphere and for the angular radius of the shadow in a special Vaidya space-time. The Vaidya metric describes a spherically symmetric object that gains or loses mass, depending on a mass function $m(v)$ that can be freely chosen. Here we consider the case that $m(v)$ is a linearly increasing or decreasing function. The first case can serve as a simple model for an accreting black hole, the second case for a (Hawking) radiating black hole. With a linear mass function the Vaidya metric admits a conformal Killing vector field which, together with the spherical symmetry, gives us enough constants of motion for analytically calculating the light-like geodesics. Both in the accreting and in the radiating case, we first calculate the light-like geodesics, the photon sphere, the angular radius of the shadow, and the red-shift of light in coordinates in which the metric is manifestly conformally static, then we analyze the photon sphere and the shadow in the original Eddington-Finkelstein-like Vaidya coordinates.
\end{abstract}


\maketitle

\section{Introduction}
\label{sec1}
As light cannot escape from black holes, the only possible way of observing them is to study their influence on light or matter in their neighborhood. When a light ray passes near a black hole, it can be deflected so strongly that it travels in a circular orbit. In the case of a spherically symmetric black hole these circular light rays fill a sphere around the black hole which is known as a \emph{photon sphere}. An observable feature, intimately related to the existence of photon spheres, is the so-called \emph{shadow} of a black hole. A major breakthrough in view of observing black holes was obtained when the Event Horizon Telescope Collaboration published an image of a black hole in 2019 \cite{article1,article2,article3,article4,article5,article6}. It shows a black disk at the center which is interpreted as the shadow. The shadow is of great interest because its special features can be used for distinguishing different types of black holes from each other and black holes from other compact objects. 

On the theoretical side, what we now call the shadow was first calculated for a Schwarzschild black hole by Synge \cite{10.1093/mnras/131.3.463} and independently by Zeldovich and Novikov \cite{ZeldovichNovikov1966}. In a Schwarzschild space-time with mass parameter $m$ there is a horizon at $r=2m$ and a photon sphere at $r=3m$. An observer will see the shadow if there are light sources everywhere around the black hole but not between the observer and the black hole. Then all past-oriented light rays that start at the observer position can be divided into two classes: Light rays of the first class are deflected by the black hole and meet on their way to infinity one of the light sources, so we associate brightness with their initial directions. Light rays of the second class go to the horizon without meeting one of the light sources, so we associate darkness with them. The borderline case consists of light rays that asymptotically spiral towards a circular light ray in the photon sphere. So the observer will see the shadow in the sky as a black circular disk whose angular radius corresponds to the angle between light rays that spiral towards the photon sphere and the radial direction. Neither Synge nor Zeldovich and Novikov used the word ``shadow'' which became common only later. Synge called the complement of the shadow the \emph{escape cone of light}; this term is still sometimes used today. 

Based on the work by Synge or Zeldovich and Novikov, it is easy to generalize the calculation of the shadow to an arbitrary spherically symmetric and static metric. This was first worked out by Pande and Durgapal \cite{PandeDurgapal1986}. The possibility of giving an analytical formula for the angular radius of the shadow relies on the fact that in a spherically symmetric and static space-time the equation for light-like geodesics is completely integrable, i.e., that there are enough constants of motion to reduce the geodesic equation to first-order form. All one has to do is to calculate the constants of motion for the circular light rays; as any light ray that asymptotically spirals against such a circular light ray must have the same constants of motion, this allows to analytically determine the boundary of the shadow in the observer's sky.  Here it is important that a light ray can spiral towards a circular light-like geodesic only if the latter is unstable with respect to radial perturbations. Therefore, the determination of unstable photon spheres is of crucial relevance for determining the shadow. For general properties of photon spheres in spherically symmetric and static space-times we refer to Claudel et al. \cite{ClaudelEtAl2001}. 

The situation is more complicated in space-times with less symmetries. In the case of a rotating object the space-time is only axially symmetric and stationary, so the equation for light-like geodesics is not in general completely integrable because there is one constant of motion less than required. However, in some axially symmetric and static space-times there is another constant of motion, known as a \emph{Carter constant}, which is not related to a Killing vector field of the space-time metric. When a Carter constant exists, the shadow can be analytically calculated. For a rotating object the shadow is flattened on one side due to the dragging of light-like geodesics by the rotating source. The most important example of a space-time that admits a Carter constant is the Kerr metric. For a stationary observer at infinity, the shape of the shadow of a Kerr black hole was first calculated by Bardeen \cite{Bardeen1973TimelikeAN}. There are a few other axially symmetric and stationary space-times for which the shadow can be analytically calculated because they admit a Carter constant. For reviews on black hole shadows, in particular in axially symmetric and stationary space-times,  we refer to Cunha and Herdeiro \cite{CunhaHerdeiro2018} and to Perlick and Tsupko \cite{perlick2021calculating}; related ray-tracing methods are detailed e.g. in a book by Zink \cite{Zink2008}.

In this paper we want to consider a different generalization of spherically symmetric and static space-times: We want to keep spherical symmetry but drop the assumption of time-independence. More specifically, we want to calculate the shadow of an accreting or radiating black hole. As spherical symmetry alone does not give us enough constants of motion for complete integrability, we will consider a special case where an additional symmetry exists. The space-time around a spherically symmetric body that gains or loses mass can be described by the Vaidya metric \cite{Vaidya1951}. This is a solution to Einstein's field equation with a null dust as the source. This null dust is either radially ingoing, causing an increase of mass, or radially outgoing, causing a decrease of mass. The dependence of the mass on time is coded in a mass function $m(v)$ that can be freely chosen. In this paper we will restrict ourselves to the case that $m(v)$ is a \emph{linearly} increasing or decreasing function. The first case gives us an idealized model for an accreting black hole, the second case for a (Hawking) radiating black hole. If $m(v)$ is linear, the space-time admits a conformal Killing vector field (cf. Nielsen \cite{Nielsen2014}) which, together with the spherical symmetry, gives us enough constants of motion for complete integrability of the equation for \emph{light-like} geodesics. This allows us to analytically calculate the (time-dependent) area of the photon sphere and the angular radius of the shadow for an accreting and for a radiating black hole.     

For related material we refer to Example 10 in Claudel et al. \cite{ClaudelEtAl2001} where the photon sphere, but not the shadow, in a piecewise defined Vaidya metric is considered. The photon sphere and the shadow in time-dependent space-times was also discussed in a recent paper by Mishra, Chakraborty and Sarkar \cite{2019}. Their study includes the Vaidya metric as a special example. However, they did not consider the particular Vaidya metrics for which the problem can be solved analytically. To the best of our knowledge, there are only two previous works where an exact analytical formula for the angular radius of the shadow was derived in a time-dependent situation: Schneider and Perlick \cite{2018} calculated the shadow of a collapsing dark star, and Perlick, Tsupko and Bisnovatyi-Kogan \cite{PhysRevD.97.104062} calculated the shadow in an expanding universe where the expansion is driven by a cosmological constant alone. There are, however, several papers where the visual appearance of a collapsing star was calculated without giving an analytical formula of the shadow: Based on the pioneering work by Ames and Thorne \cite{AmesThorne1968} several authors calculated the redshift of the surface of a collapsing star. More recent papers by Kong et al. \cite{KongMalafarinaBambi2014,KongMalafarinaBambi2015} and by Ortiz et al. \cite{OrtizSarbachZannias2015a,OrtizSarbachZannias2015b} investigated the frequency shift of light passing through a collapsing transparent star, thereby contrasting the collapse to a black hole with the collapse to a naked singularity. Moreover, we also want to direct the reader's attention to a very recent paper by Koga et al. \cite{KogaEtAl2022} where the shadow of an accreting black hole is calculated numerically.

We organize the paper as follows. In section \ref{sec2}, we investigate the evolution of the photon sphere and the shadow of a Vaidya black hole with linearly increasing mass function. In subsection \ref{subsec2A}, it is our goal to determine the light-like geodesics in this space-time where, because of the spherical symmetry, we may restrict ourselves to the equatorial plane. To that end we perform a specific coordinate transformation that makes the space-time manifestly conformally static and allows us to analytically calculate the light-like geodesics. Based on these results, we determine in subsection \ref{subsec2B} the location of the photon sphere in the conformally static coordinates. In subsection \ref{subsec2C}, we find an analytical formula for the angular radius of the shadow as seen by a conformally static observer, i.e., by an observer whose worldline is an integral curve of the conformal Killing vector field. We calculate the red-shift under which one conformally static observer sees another one in subsection \ref{subsec2D}. Finally, in the last subsection \ref{subsec2E}, we analyze the results in the original coordinates. In particular, we calculate the angular radius of the shadow as seen by an observer adapted to the original coordinates; with respect to the conformally static observers, such an observer is moving towards the black hole. -- In section \ref{sec3} we proceed along the same lines for the black hole with linearly decreasing mass function. Finally, in the discussion section \ref{sec4}, we summarize the results obtained in this paper.

\section{Photon sphere and shadow of a black hole with increasing mass}
\label{sec2}
The Vaidya metric \cite{Vaidya1951} describes spherically symmetric (i.e., non-rotating) stars or black holes, emitting or absorbing a null dust. Thus, their mass will be accordingly decreasing or increasing, in contrast to the Schwarzschild space-time where the mass is a constant. As a consequence, the Vaidya metric describes a dynamical space-time instead of a static space-time as the Schwarzschild metric does. In the real world, astronomical bodies gain mass when they absorb radiation and they lose mass when they emit radiation which means that the space-time around them is time-dependent. The Vaidya metric provides a comparatively simple but interesting setting for studying the properties of such dynamical space-times, in particular the existence of a photon sphere and the formation of a shadow.

In Eddington-Finkelstein-like coordinates, the Vaidya metric is given by the equation 
\begin{equation}
\label{1}
    ds^2 = -\left( 1 - \frac{2 m(v) }{r} \right)dv^2 \pm 2 dv dr + r^2(d\theta^2 + \mathrm{sin} ^2\theta \, d\phi^2) \, .
\end{equation}
For constant $m(v)$ this reduces to the Schwarzschild metric. For non-constant $m(v)$ the metric describes the gravitational field around a central object with increasing or decreasing mass, where the mass depends on the coordinate $v$. Note that the hypersurfaces $v = \mathrm{const.}$ are spanned by the light-like vector field $\partial _r$ and the space-like vector fields $\partial _{\theta}$ and $\partial _{\phi}$, so these hypersurfaces are light-like. If the metric ({1}) is inserted on the left-hand side into Einstein's field equation, the right-hand side gives us an energy-momentum tensor of the form
\begin{equation}
T^{\rho \nu} = \pm \dfrac{8 \pi G}{c^4} \, m'(v) \, K^{\rho} K^{\nu} \, , \qquad 
K^{\rho} \partial _{\rho} = \mp \, \partial _r \, .
\label{eq:Tgen}
\end{equation}
This is the energy-momentum tensor of a null dust. In (\ref{1}) and (\ref{eq:Tgen}) we have to choose always the upper sign or always the lower sign. Note that then the vector field $K^{\rho} \partial _{\rho}$ is always future-oriented, assuming that the time-orientation has been chosen such that the coordinate $v$ is increasing towards the future. If we choose the upper sign, the coordinates are referred to as ``ingoing Eddington-Finkelstein-like coordinates''. Then the null dust is moving in the direction of decreasing $r$ and the energy-density is positive if $m'(v)>0$ which means that the central object is accreting by absorbing infalling matter of positive energy density. This is the case we will consider in this section. More specifically, we will consider the special case that $m(v)$ is a \emph{linearly} increasing function, $m(v) = \mu v$ with a positive constant $\mu$. Then the metric reads 
\begin{equation}
\label{2}
    ds^2 = -\left( 1 - \frac{2 \mu v }{r} \right)dv^2 + 2 dv dr + r^2(d\theta^2 + \mathrm{sin} ^2 \theta \, d\phi^2)
\end{equation}
and, by Einstein's field equation, 
\begin{equation}
T^{\rho \nu} = \dfrac{8 \pi G}{c^4} \, \mu  \, K^{\rho} K^{\nu} \, , \qquad 
K^{\rho} \partial _{\rho} = - \, \partial _r \, .
\label{eq:Taccr}
\end{equation}

We will see that for $0< 16 \mu<1$ the space-time features two horizons; it can then be interpreted as a simple and idealized but useful model for an accreting \emph{black hole}. The coordinates $\theta$ and $\phi$ have their usual range on the sphere, whereas $-\infty < v < \infty$ and $0 < r < \infty$. On this domain the metric (\ref{2}) is obviously regular. As the Kretschmann scalar equals $48 \mu^2 v^2 /r^6$, there is a curvature singularity at $r=0$, i.e., it is impossible to extend the metric beyond this range. On the domain $- \infty < v < 0$ the mass is negative which should be considered as unphysical. Therefore, we restrict our consideration to the domain $0 < v < \infty$ and we assume that to the past of this domain the space-time is given by some other metric. The latter has not to be specified because it is irrelevant for the construction of the shadow, as we will see below.  One possibility is to assume that  $m(v) = 0$ (Minkowski space-time) for $v<0$. Such models, where a singularity forms at $v=0$, have been considered first by Kuroda \cite{Kuroda1984b} and by Papapetrou \cite{Papapetrou1986} and are sometimes called \emph{Vaidya-Kuroda-Papapetrou} models. For a more detailed discussion of Vaidya metrics and the nature of their horizons and singularities we refer the reader to Section 9.5 of the book by Griffiths and Podolsk{\'y} \cite{griffiths_podolsky_2009}.

The dimensionless number $\mu$ is a measure of the accretion rate of the black hole. An observer on a $v$-line assigns the mass $m(v_0)$ to the black hole when crossing the hypersurface  $v = v_0$. Here and in the following, by an ``observer on a $v$-line'' we mean an observer whose worldline has constant $r$, $\theta$ and $\phi$ coordinates.
From the metric we read that such worldlines are indeed time-like if $r > 2 \mu v$ and that proper time $\tau$ is related to the coordinate $v$ by $c^2 d \tau ^2 = (1- 2 \mu v / r ) dv^2$ where $c$ is the vacuum speed of light. This demonstrates that $v/c$ can be identified with proper time if the observer is far away from the black hole. As in SI units the mass of the black hole is $M = c^2 \mu v /G$, the accretion rate, as seen by such a distant observer, is 
\begin{equation}
    \dfrac{dM}{d \tau} \approx c \dfrac{d M}{dv} = c^3 \mu /G \, .
\end{equation}
Here $G$ is Newton's gravitational constant.
For a very rough order-of-magnitude estimate of the highest value of $\mu$ that can be expected in an astrophysical situation, we may consider a supermassive black hole of $10^{10} M_{\odot}$ and an accretion rate that exceeds the Eddington accretion rate by two orders of magnitude; this is not totally unrealistic, see e.g. Pu et al. \cite{SEAMBH2015}. This would give an accretion rate of 
\begin{equation}
    \dfrac{dM}{d \tau} \approx 10^3 M_{\odot}/\mathrm{yr} \, , \quad \mu \approx 10^{-8} \, .
\label{eq:muest}
\end{equation}

\subsection{Light-like geodesics in the equatorial plane}\label{subsec2A}

To understand the behavior of light traveling in the metric (\ref{2}), we formulate the equations of motion for light-like geodesics in this metric. The metric (\ref{2}) is obviously spherically symmetric, but this symmetry alone does not provide us with sufficiently many constants of motion for reducing the geodesic equation to first order. Apparently there is no additional symmetry. Actually, the general Vaidya metric (\ref{1}) does not have any additional symmetry. The more special metric (\ref{2}), however, admits an additional conformal Killing vector field, as was already discussed by Nielsen \cite{Nielsen2014}, which allows us to analytically determine the \emph{light-like} geodesics. We mention that, more precisely, a Vaidya metric (\ref{1}) admits an additional conformal Killing vector field \emph{only} if $m(v)$ is a linear function. This follows from the work of Ojako et al. \cite{OjakoEtAl2020} who determined the existence of a conformal Killing vector field in a class of space-times that includes the traditional Vaidya metrics. As we are free to add a constant to the coordinate $v$, (\ref{2}) is indeed the most general form of a Vaidya metric with increasing mass function where an additional  conformal Killing vector field exists. To make the existence of this conformal Killing vector field for the metric (\ref{2}) explicit, we perform the coordinate transformation
\begin{equation}
\label{3}
    v = r_0 e^{cT/r_0} \quad\mathrm{and}\quad r = Re^{cT/r_0}
\end{equation}
where $ r_0 $ is a positive constant with the dimension of a length. In the new coordinates the metric (\ref{2}) takes the following form:
\begin{equation}
\label{4}
     ds^2 = e^{2cT/r_0}\left\{-\left( 1 - \frac{2 \mu r_0 }{R} - \frac{2R}{r_0} \right)c^2 dT^2 + 2 c dT dR + R^2(d\theta^2 + \mathrm{sin} ^2\theta d\phi^2)\right\} \, .
\end{equation}
We read that in this representation the metric is regular on the domain $ -\infty < T < \infty $ and $ 0 < R < \infty $. By (\ref{3}), this corresponds to the domain $ 0 < v < \infty $ and $ 0 < r < \infty $ in the original coordinates.  This is precisely the domain to which we want to restrict our consideration.

From (\ref{4}) it is obvious that $\partial / \partial T$ is a conformal
Killing vector field. This vector field is time-like if 
\begin{equation}
\label{5}
    1 - \frac{2 \mu r_0 }{R} - \frac{2R}{r_0} > 0 ,
\end{equation}
i.e., on the domain 
\begin{equation}
\label{6}
    R_- < R < R_+ 
\end{equation}
where 
\begin{equation}
\label{7}
R_{\pm} = \frac{r_0}{4}(1 \pm \sqrt{1 - 16\mu}) 
\end{equation}
are the positions of two horizons. Obviously these are
conformal Killing horizons. In the following we assume that $ 16 \mu < 1$ to make sure that we have a region where $\partial /\partial T$ is time-like. The inner horizon at $R_-$ is then to be interpreted as a black hole horizon.

We will now determine the light-like geodesics in the metric (\ref{4}). Later we will then transform back the results to the original coordinates $(v,r,\theta,\phi)$. The light-like geodesics in the equatorial plane obey the following equation of motion derived from equation (\ref{4})
\begin{equation}
\label{8}
   0 =  -\left( 1 - \frac{2 \mu r_0 }{R} - \frac{2R}{r_0} \right)c^2 \dot{T}^2 + 2 c \dot{T}\dot{R} + R^2\dot{\phi}^2 \, .
\end{equation}
The crucial observation is that $\partial / \partial T$ is a conformal Killing vector field. This gives us along every light-like geodesic the following constant of motion $E$:
\begin{equation}
\label{9}
    -\left( 1 - \frac{2 \mu r_0 }{R} - \frac{2R}{r_0} \right) e^{2cT/r_0} c \dot{T}  + e^{2cT/r_0} \dot{R} = E \, .
\end{equation}
Another constant of motion $L$ derives from the fact that $\partial / \partial \phi$ is a Killing vector field:
\begin{equation}
\label{10}
e^{2cT/r_0}R^2 \dot{\phi} = L \, .
\end{equation}

\begin{figure*}[ht!]
\includegraphics[width = .65\textwidth]{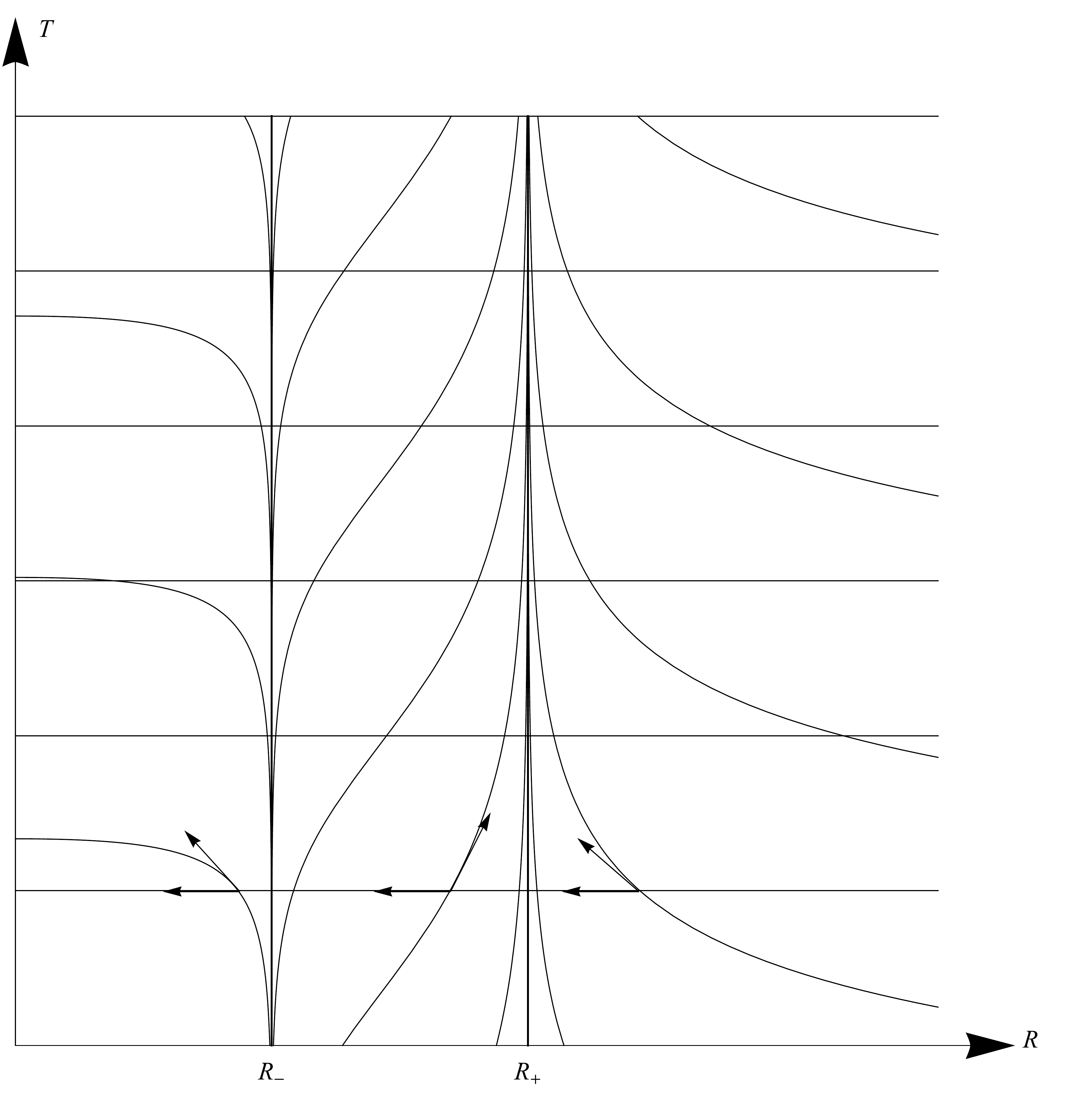}
\caption{{Radial light rays}}
\label{fig:radial}
\end{figure*}

From (\ref{8}) we read that \emph{radial} light rays ($\dot{\phi}=0$) satisfy \begin{equation}
    c \, dT =0 \, , \quad c \, dT = \frac{2dR}{1 - \frac{2 \mu r_0 }{R} - \frac{2R}{r_0}} 
\label{eq:dradial}
\end{equation}
where the first equation is for ingoing and the second for outgoing radial light rays. Integration of these equations results in
\begin{equation}
    c \, T = \mathrm{const.} \, , \quad c \, T = \frac{r_0}{R_+ - R_-}\left\{ R_- \mathrm{log} \left(\frac{R}{R_-}-1\right) -  R_+ \mathrm{log} \left(1-\frac{R}{R_+}\right) \right\} + \mathrm{const.}
\label{eq:radial}
\end{equation}
These radial light rays are plotted in Fig. \ref{fig:radial}. The little arrows indicate the future direction.

For non-radial light rays we divide equations (\ref{8}) and (\ref{9}) by $ \dot{\phi} ^2 $ and $\dot{\phi}$, respectively, and use equation (\ref{10}) to find
\begin{equation}
\label{11}
    -\left( 1 - \frac{2 \mu r_0 }{R} - \frac{2R}{r_0} \right)c^2 \left(\frac{dT}{d\phi}\right)^2 + 2 c \frac{dT}{d\phi} \frac{dR}{d\phi} + R^2 = 0 \, ,
\end{equation}
\begin{equation}
\label{12}
    -\left( 1 - \frac{2 \mu r_0 }{R} - \frac{2R}{r_0} \right)c \frac{dT}{d\phi} + \frac{dR}{d\phi} = \frac{E R^2}{L} \, .
\end{equation}
By solving equations (\ref{11}) and (\ref{12}) for $ \frac{dR}{d\phi} $ and $ \frac{dT}{d\phi} $, we get the following equations describing light-like geodesics in the equatorial plane for the metric (\ref{4}):
\begin{equation}
\label{13}
    \frac{dR}{d\phi} = \pm  \sqrt{\frac{E^2 R^4}{L^2} - R^2 + 2\mu r_0 R + \frac{2R^3}{r_0}} \, ,
\end{equation}
\begin{equation}
\label{14}
    c\frac{dT}{d\phi} = \frac{-\frac{E R^2}{L} \pm  \sqrt{\frac{E^2 R^4}{L^2} - R^2 + 2\mu r_0 R + \frac{2R^3}{r_0}}}{1 - \frac{2 \mu r_0 }{R} - \frac{2R}{r_0}} \, .
\end{equation}
Eq. (\ref{13}) determines $R$ as a function of $\phi$. Thereupon, (\ref{14}) can be used to determine $T$ as a function of $\phi$.

\subsection{The photon sphere}
\label{subsec2B}
We will now show with the help of (\ref{13}) that the metric (\ref{4}) with $\mu < 1/16$ admits exactly one photon sphere, and that it is located between the two horizons. We first evaluate the  condition $ \frac{dR}{d\phi} = 0 $ which will give us the extremum points of light paths, $ R = R_m $. In a second step we will then determine the radius coordinate $R_p$ of the photon sphere by requiring in addition that $ \frac{d^2R}{d\phi^2} = 0$ at $R_p$. By (\ref{13}) the condition $ \frac{dR}{d\phi} = 0$ holds at $R=R_m$ if and only if
\begin{equation}
\label{15}
    \frac{E^2 R_m^3}{L^2} - R_m + 2\mu r_0 +  \frac{2R_m^2}{r_0} = 0 \, .
\end{equation}
From equation (\ref{15}), we can write $ \frac{E^2}{L^2} $ in terms of $R_m$ as follows,
\begin{equation}
\label{16}
    \frac{E^2}{L^2} = \frac{R_m - 2\mu r_0 - \frac{2 R_m^2}{r_0}}{R_m^3} \, .
\end{equation}
Now we compute $\frac{d^2 R}{d\phi^2}$ at $ R = R_m $ from (\ref{13}):
\begin{equation}
\label{17}
    \frac{d^2R}{d\phi^2}\bigg|_{R=R_m} = \frac{4E^2R_m^3}{L^2} - 2R_m + 2\mu r_0 + \frac{6R_m^2}{r_0} \, .
\end{equation}
By (\ref{16}), eq. (\ref{17}) results in
\begin{equation}
\label{18}
    \frac{d^2R}{d\phi^2}\bigg|_{R=R_m} = -\frac{2}{r_0}(R_m^2 - r_0 R_m + 3 \mu r_0^2) 
\end{equation}
which can be rewritten as 
\begin{equation}
\label{19}
    \frac{d^2R}{d\phi^2}\bigg|_{R=R_m} = -\frac{2}{r_0}\left\{ R_m - \frac{r_0}{2}(1 - \sqrt{1 - 12\mu})  \right\}\left\{ R_m - \frac{r_0}{2} (1 + \sqrt{1 - 12\mu})   \right\} \, .
\end{equation}
On the other hand, we can rewrite eq. (\ref{15}) in the following form:
\begin{equation}
\label{20}
    \frac{E^2}{L^2} = \frac{2 \left( R_m - \frac{r_0}{4}   (1 - \sqrt{1 - 16\mu}\right) \left(  \frac{r_0}{4}(1 + \sqrt{1 - 16\mu}) - R_m \right)}{R_m^3} \, .
\end{equation}
Comparison of this equation with (\ref{7}) shows that $  \frac{E^2}{L^2} > 0 $ only for $ R_- < R_m < R_+$, i.e., turning points occur only in the region between the two horizons where the conformal Killing vector field $\partial / \partial T$ is time-like. Thus, light travelling inside of both the horizons will move towards the central singularity without any extremum point and light travelling outside of both the horizons will move towards infinity without having any extremum point. 
\par This observation implies that a photon sphere can exist only in the region between the two horizons. To find the exact location of a photon sphere we need to find a value $R_m$ with $ R_- < R_m < R_+$ such that the right-hand side of (\ref{19}) vanishes. The only possible solution of this problem is the following value of $ R_m=R_p $:
\begin{equation}
\label{21}
    R_p = \frac{r_0}{2}(1 - \sqrt{1 - 12\mu}) \, .
\end{equation}

\begin{figure*}[ht!]
\includegraphics[width = .45\textwidth]{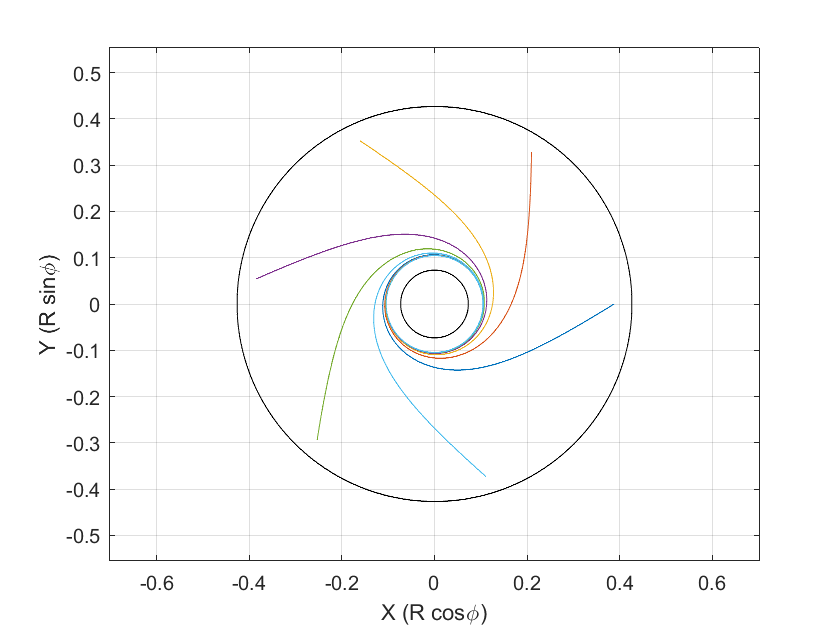}
\caption{{Light rays spiraling towards the photon sphere ($ \mu = 1/32 $)}}
\label{fig1}
\end{figure*}

Thus, there is one photon sphere, located at the radius $ R_p$ given by (\ref{21}), for the metric (\ref{4}) with $ 0 < 16 \mu < 1 $ as shown in Fig. \ref{fig1}.  In this figure, the two horizons are shown as black circles and lengths are given in units of $r_0$.

For $R_m=R_p$ we find from (\ref{16}) that
\begin{equation}
\label{22}
    \dfrac{E}{L} = \dfrac{\pm  \sqrt{ 1 - \dfrac{2 \mu r_0}{R_p}-2 \dfrac{R_p}{r_0} }}{R_p} \, ,
\end{equation}
so (\ref{14}) gives us the following relation between $T$ and $\phi$ for an equatorial light-like geodesic in the photon sphere :
\begin{equation}
\label{23}
c T = \frac{\pm \, R_p \, \phi}{\sqrt{1 - \frac{2 \mu r_0 }{R_p} - \frac{2R_p}{r_0}}} \, . 
\end{equation}
As $T$ runs over all of $\mathbb{R}$, so does $\phi$. By (\ref{10}), this implies that the light-like geodesics in the photon sphere are non-extendable: They start at the singularity and the affine parameter then runs up to $\infty$. The latter remark is important, in particular in view of the fact that the coordinates $(T,R,\phi,\theta)$ cover only half of the space-time.

Light-like geodesics which have a ratio of constants $ \frac{E}{L} $ given by equation (\ref{22}) are either completely contained in the photon sphere or they spiral asymptotically towards the photon sphere as shown in Fig. \ref{fig1}. Here it is important to realize that, although the metric (\ref{4}) depends on $T$, the photon sphere and the two horizons exist at \emph{fixed} $R$-coordinates for all $T$. Also note that the area of the photon sphere is  \emph{not} equal to $A = 4 \pi R_p ^2$ with $R_p$ from (\ref{21}); from (\ref{4}) we read that this area rather depends on $T$ and is equal to 
\begin{equation}
\label{24}
    A= 4 \pi r_p ^2=4 \pi e^{2cT/r_0} R_p^2 = \pi e^{2cT/r_0} r_0^2 \big( 1 - \sqrt{1 - 12\mu} \, \big)^2 \, ,
\end{equation}
so it goes to 0 for $T \to - \infty$ and to $\infty$ for $T \to \infty$. Again according to (\ref{4}), along a $T$-line at the photon sphere proper time $\tau$ is related to $T$ by 
\begin{equation}
    d \tau  = e^{c T /r_0} \sqrt{1- \dfrac{2 \mu r_0}{R_p} - \dfrac{2 R_p}{r_0}} \, d T \, . 
\end{equation}
This implies that the area radius $r_p$ of the photon sphere increases with proper time at a constant rate,
\begin{equation}
    \dfrac{d r_p}{d \tau} = \dfrac{R_p c}{r_0 \sqrt{1- \dfrac{2 \mu r_0}{R_p} - \dfrac{2 R_p}{r_0}}} = \sqrt{27} \, \mu \, c \Big( 1+ O \big( \mu  \big) \Big) \, .
\end{equation}
For the accretion rate according to (\ref{eq:muest}) this gives a velocity of approximately $10^{-7} c$ for the expansion of the photon sphere.

\subsection{The angular radius of the shadow}
\label{subsec2C}
\begin{figure*}[ht!]
\includegraphics[width = .5\textwidth]{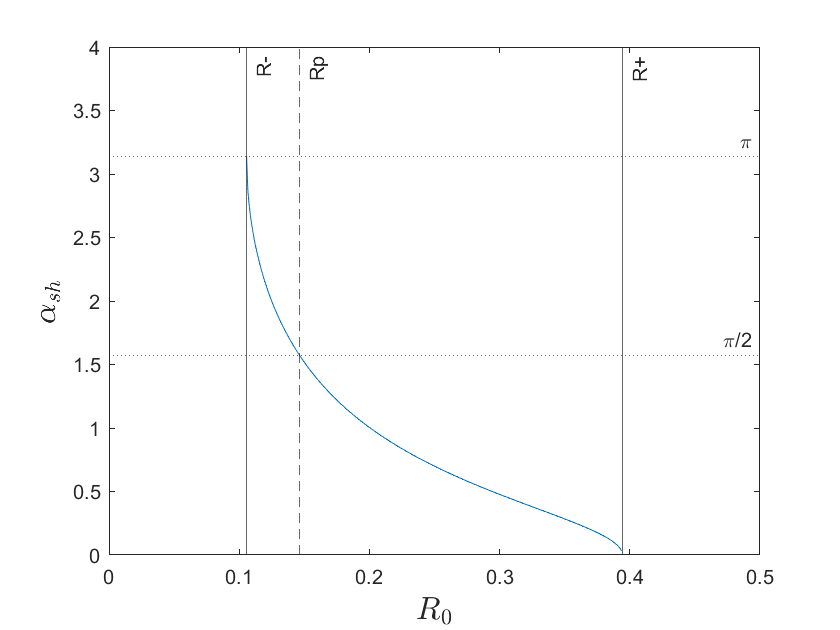}
\caption{Angular radius of the shadow ($ \mu = 1/24 $) }
\label{fig2}
\end{figure*}

Here we find the angular radius of the black hole shadow seen  by an observer on a $T$-line in the region between the two horizons. We will see that this angular radius is time-independent, although the area of the photon sphere increases with $T$. To find the angular radius of the shadow, we use the following tetrad field for $  R_- < R < R_+ $.
\begin{equation}
\label{25}
    e_0 = \frac{e^{-cT/r_0}}{\left(\sqrt{ 1 - \frac{2 \mu r_0 }{R} - \frac{2R}{r_0}}\right)c}\frac{\partial}{\partial T} \, ,
\end{equation}
\begin{equation}
\label{26}
    e_1 = \frac{e^{-cT/r_0}}{\left(\sqrt{ 1 - \frac{2 \mu r_0 }{R} - \frac{2R}{r_0}}\right)c}\frac{\partial}{\partial T} + \left(\sqrt{ 1 - \frac{2 \mu r_0 }{R} - \frac{2R}{r_0}}\right)e^{-cT/r_0}\frac{\partial}{\partial R} \, ,
\end{equation}
\begin{equation}
\label{27}
    e_2 = \frac{e^{-cT/r_0}}{R}\frac{\partial}{\partial \theta} \quad\mathrm{and}\quad e_3 = \frac{e^{-cT/r_0}}{R \, \mathrm{sin} \, \theta}\frac{\partial}{\partial \phi} \, .
\end{equation}
In the equatorial plane we consider a light-like geodesic $ (T(\lambda), R(\lambda), \phi(\lambda)) $, where $ \lambda $ is an affine parameter. We then expand the light-like geodesic's tangent vector with respect to the tetrad (\ref{25}) to (\ref{27}). Since the tangent vector is light-like, the expansion can be written in terms of an angle $ \alpha $  as follows:
\begin{equation}
\label{28}
    \dot{T}\frac{\partial}{\partial T} + \dot{R}\frac{\partial}{\partial R} + \dot{\phi}\frac{\partial}{\partial \phi} = \xi \left( e_0 + e_1 \mathrm{cos} \, \alpha - e_3 \mathrm{sin} \, \alpha \right)
\end{equation}
where $ \xi $ is a scalar factor and $ \alpha $ is the angle between the light-like geodesic and the radial direction in the rest system of the observers governed by the tetrad (\ref{25}) to (\ref{27}). Comparing the coefficients of $\frac{\partial}{\partial R}$ and $\frac{\partial}{\partial \phi}$ in (\ref{28}) yields
\begin{equation}
\label{29}
    \dot{R} = \mathrm{cos} \, \alpha \left(\sqrt{ 1 - \frac{2 \mu r_0 }{R} - \frac{2R}{r_0}}\right)e^{-cT/r_0} \, ,
\end{equation}
\begin{equation}
\label{30}
    \dot{\phi} = -\frac{(\mathrm{sin} \, \alpha) e^{-cT/r_0}}{R} \, .
\end{equation}
From equations (\ref{13}), (\ref{29}) and (\ref{30}), we have
\begin{equation}
\label{31}
    \frac{1}{\frac{E^2 R^4}{L^2} - R^2 + 2\mu r_0 R + \frac{2R^3}{r_0}} = \frac{\mathrm{sin}^2\alpha}{\mathrm{cos}^2\alpha\left( 1 - \frac{2 \mu r_0 }{R} - \frac{2R}{r_0}\right)R^2} \, .
\end{equation}
From equations (\ref{16}) and (\ref{31}), we can write $ \mathrm{sin} \, \alpha $ in terms of $ R_m $ as follows:
\begin{equation}
\label{32}
    \mathrm{sin} \, \alpha = \sqrt{\frac{R_m^3\left(R - 2\mu r_0 - \frac{2R^2}{r_0}\right)}{R^3\left(R_m - 2\mu r_0 - \frac{2R_m^2}{r_0}\right)}} \, .
\end{equation}
Now the photon sphere is at the radius $ R_p = \frac{r_0}{2}(1 - \sqrt{1 - 12\mu}) $. Thus, the angular radius of the shadow of the photon sphere for an observer at $ R = R_0 $ becomes
\begin{equation}
\label{33}
    \mathrm{sin} \, \alpha_{sh} = \sqrt{\frac{R_p^3\left(R_0 - 2\mu r_0 - \frac{2R_0^2}{r_0}\right)}{R_0^3\left(R_p - 2\mu r_0 - \frac{2 R_p^2}{r_0}\right)}} = \sqrt{\frac{R_p^3 (R_0-R_-) (R_+-R_0)}{R_0^3 (R_p-R_-) (R_+-R_p)}} \, .
\end{equation}
Of course, the expression under the square-root must be between 0 and 1 which is indeed the case for $ R_- < R_0 < R_+ $, i.e., in the region where the conformally static observers exist. Eq. (\ref{33}) must be supplemented with the information that $\alpha _{\mathrm{sh}}$ is in the interval between 0 and $\pi/2$ for $R_p < R_0<R_+$ and in the interval between $\pi/2$ and $\pi$ for $R_-<R_0<R_p$. We have $\alpha _{\mathrm{sh}} \to 0$ (bright sky) for $R_0 \to R_+$ and $\alpha _{\mathrm{sh}} \to \pi$ (dark sky) for $R_0 \to R_-$. At $R_0=R_p$ the shadow covers half of the sky which is obvious without any calculation. In Fig. \ref{fig2}, we have plotted the angular radius of the black hole shadow $ \alpha_{sh} $ as a function of $ R_0 $. The solid black lines labeled $ R_- $ and $R_+$ mark the  inner and outer horizon, respectively, and the dashed black line labeled $ R_p $ represents the photon sphere. $R_0$ is given in units of $r_0$.

As an alternative, the shadow formula (\ref{33}) can be rewritten as
\begin{equation}
    \mathrm{cos} \, \alpha_{sh} = \dfrac{(R_0-R_p)}{R_0} \sqrt{1+ \frac{2 \mu r_0 R_p}{R_0 (4 \mu r_0 - R_p)}} \, .
\label{eq:shadowstat}
\end{equation}
This version gives a unique value for $\alpha _{\mathrm{sh}}$ in the interval between 0 and $\pi$, for all $R_-<R_0<R_+$.

With the shadow known for a conformally static observer, the shadow can be calculated for an observer in arbitrary motion with the help of the aberration formula. We will exemplify this in Section \ref{subsec2E} below. As the aberration formula maps circles in the sky onto circles in the sky, \emph{any} observer in the region between the two horizons sees a circular shadow.

\subsection{The red-shift of light}
\label{subsec2D}
In this section we compute the red-shift in the metric (\ref{4}). We assume two observers, labeled $A$ and $B$ respectively, who are both moving on $T$-lines. This is of course only possible in the region between the two horizons. The red-shift $ z =  \frac{\nu_A}{\nu_B} - 1$ is determined by the frequency-ratio
\begin{equation}
\label{34}
    \frac{\nu_A}{\nu_B} = \sqrt{\frac{g_{00}(T_B,R_B)}{g_{00}(T_A,R_A)}}
\end{equation}
where $ \nu_A $ and $ \nu_B $ are the frequency of the light at $ (T_A, R_A , \theta _A , \phi _A) $ and $ (T_B, R_B,\theta _B , \phi _B) $, respectively. From (\ref{4}), equation (\ref{34}) becomes
\begin{equation}
\label{35}
     \frac{\nu_A}{\nu_B} = \frac{\sqrt{ 1 - \frac{2 \mu r_0 }{R_B} - \frac{2R_B}{r_0}}}{\sqrt{ 1 - \frac{2 \mu r_0 }{R_A} - \frac{2R_A}{r_0}}} \, e^{c(T_B-T_A)/r_0} \, .
\end{equation}
As $\partial / \partial T$ is a conformal Killing vector field, $T_B-T_A$ has the same value for all light rays from $A$ to $B$, i.e., the red-shift is time-independent. Note that this result comes from the fact that the conformal factor in front of the metric (\ref{4}) is an exponential function of $T$. If this factor would depend on $T$ in any other form, $T_B$ and $T_A$ would enter not only in the form of their difference and the red-shift between two observers on $T$-lines would depend on $T$.   

We now specify to the case that the considered light rays travel in the radial direction which requires $(\theta _A, \phi _A)= (\theta _B, \phi _B)$. Then we find from (\ref{eq:radial}) that (\ref{35}) becomes
\begin{equation}
\label{37}
     \frac{\nu_A}{\nu_B} = \frac{\sqrt{ 1 - \frac{2 \mu r_0 }{R_B} - \frac{2R_B}{r_0}}}{\sqrt{ 1 - \frac{2 \mu r_0 }{R_A} - \frac{2R_A}{r_0}}}
\end{equation}
for ingoing and 
\begin{equation}
\label{39}
     \frac{\nu_A}{\nu_B} = \frac{\sqrt{ 1 - \frac{2 \mu r_0 }{R_B} - \frac{2R_B}{r_0}}}{\sqrt{ 1 - \frac{2 \mu r_0 }{R_A} - \frac{2R_A}{r_0}}}\left\{ \left( \frac{R_B - R_-}{R_A - R_-} \right)^{\frac{R_-}{R_+ - R_-}} \left( \frac{R_+ - R_A}{R_+ - R_B} \right)^{\frac{R_+}{R_+ - R_-}}  \right\}
\end{equation}
for outgoing light rays.

\subsection{Analysis in the original coordinates}
\label{subsec2E}

Now we transform back the obtained results to the original coordinate system $(v,r,\theta,\phi)$. This allows us to analyze the behavior of the photon sphere as seen by observers on $v$-lines (lines of constant $ r, \theta, \phi $).

From (\ref{2}) we read that the $v$-lines are time-like on the domain where
\begin{equation}
\label{40}
  2 \mu v < r \, , \quad 2 \mu r_0 < R \, .
\end{equation}
This domain includes the region between the horizons. The $v$-lines are plotted as dashed curves in Fig. \ref{fig:vt}. The domain where the $v$-lines fail to be time-like (i.e., where (\ref{40}) is violated) is shown in dark gray. The other features of Fig. \ref{fig:vt} will be explained below. We see that the observers on $v$-lines come from infinity, first cross the outer horizon at $R_+$, then the photon sphere at $R_p$ and then the inner horizon at $R_-$.

For analyzing the situation in the coordinates $(v,r,\theta,\phi)$ it is important to realize that the hypersurfaces $v = \mathrm{const.}$ are light-like everywhere and that $r$ is an \emph{area coordinate}, i.e., that the sphere $(v,r)= \mathrm{const.}$ has area $4 \pi r^2$, as can be read from (\ref{2}).

\begin{figure*}[ht!]
\includegraphics[width = .45\textwidth]{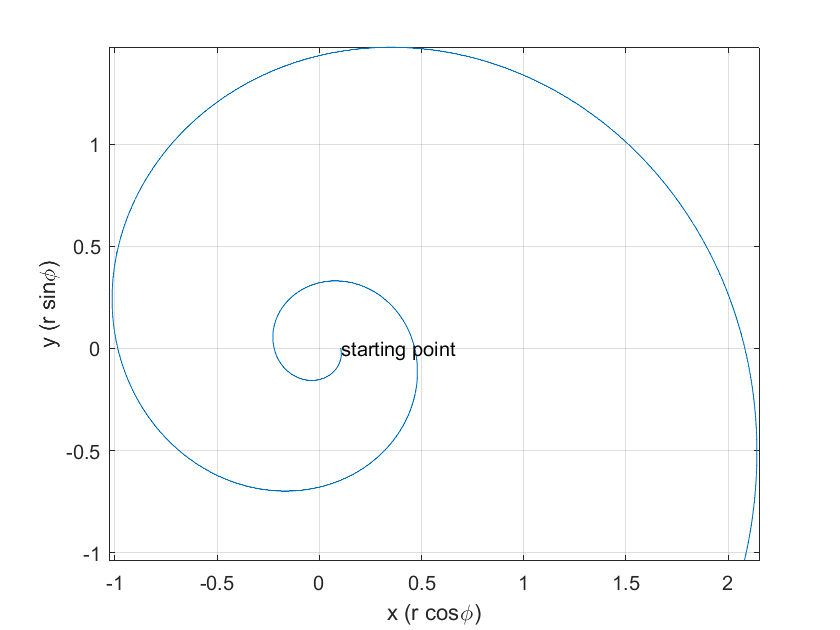} 
\caption{Light path in the photon sphere in $(v, r, \theta, \phi)$ coordinates ($ \mu = 1/24 $)}
\label{fig3}
\end{figure*}

\begin{figure*}[ht!]
\includegraphics[width = .49\textwidth]{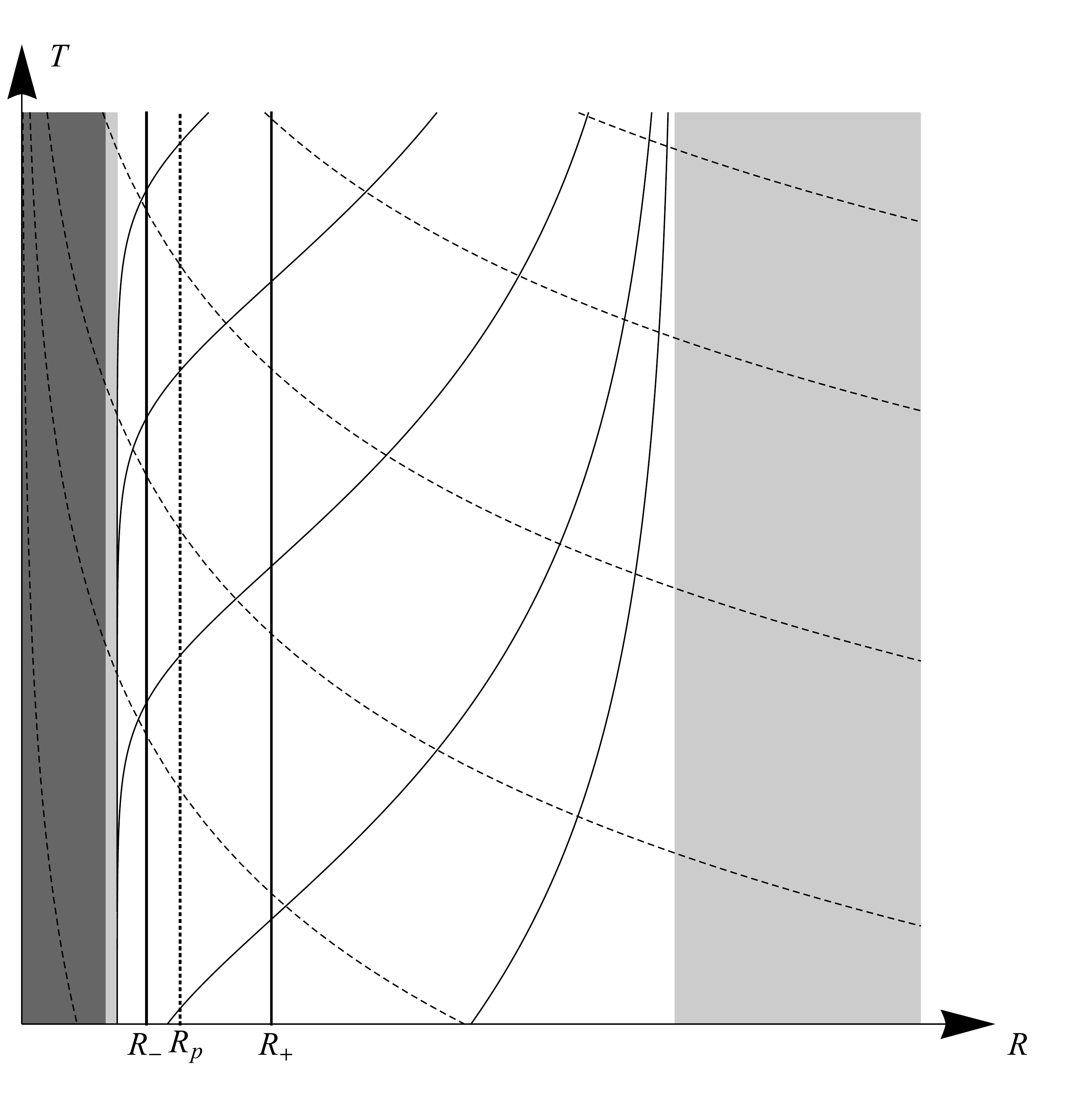} 
\includegraphics[width = .48\textwidth]{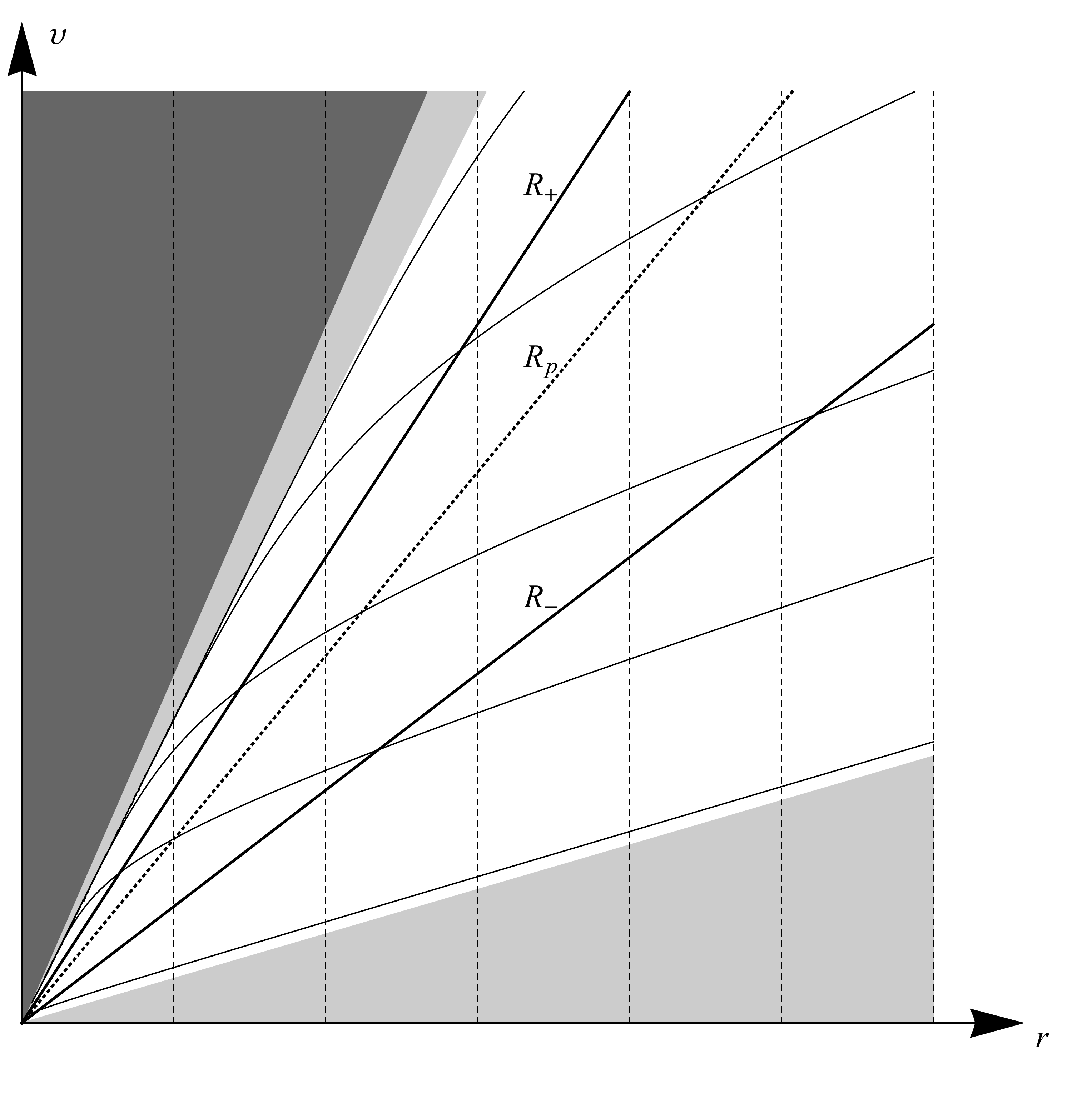} 
\caption{$v$-lines (dashed) and hypersurfaces $t = \mathrm{const.}$ (solid)}
\label{fig:vt}
\end{figure*}

We have found above that in the $ (T, R, \theta, \phi) $ coordinates light paths in the photon sphere are at $R=R_p= \frac{r_0}{2}(1 - \sqrt{1 - 12\mu}) $ and that in the equatorial plane their $T$ and $\phi$ coordinates are related by (\ref{23}). Transforming these two equations back to the original coordinates $(v, r, \theta, \phi)$ with the help of (\ref{3}) results in
\begin{equation}
\label{41}
    r = \frac{R_p v}{r_0} \quad\mathrm{and}\quad \phi =  \pm \left(\frac{r_0}{R_p}\sqrt{1 - \frac{2\mu r_0}{R_p} - \frac{2R_p}{r_0}}\right) \mathrm{log} \left( \frac{v}{r_0} \right) \, .
\end{equation}
These equations give us the path of the geodesic parametrized by $v$. We see that in these coordinates the light-like geodesics in the photon sphere spiral outwards, see Fig. \ref{fig3}, i.e., that the photon sphere expands. In this figure lengths are given in units of $r_0$. The radius coordinate $r$ is linearly increasing with $ v $. The angle $ \phi $ is a logarithmic function of $v$, so the angular speed $ \frac{d\phi}{d v} $ is inversely proportional to $ v $.

Whereas $v$ is appropriate for parametrizing individual light rays, we have to keep in mind that the hypersurfaces $v = \mathrm{const.}$ are light-like, i.e., that they cannot be interpreted as equal-time slices. For describing the expansion of the entire photon sphere, as observed by observers on $v$-lines, we have to introduce an appropriate time function. We define such a function $t$ in the $(T,R,\theta,\phi)$ coordinates by the equation
\begin{equation}
\label{42}
    c \, dt = c \, dT - \dfrac{dR}{1 - \dfrac{2 \mu r_0}{R}-\dfrac{R}{r_0}} \, .
\end{equation}
With the help of the relation
\begin{equation}
\label{43}
    \dfrac{\partial}{\partial v} = e^{-cT/r_0} \dfrac{1}{c} \Big( \dfrac{\partial}{\partial T}- \dfrac{cR}{r_0} \dfrac{\partial}{\partial R} \Big)
\end{equation}
which follows from (\ref{3}), it is easy to verify that the hypersurfaces $t = \mathrm{const.}$ are orthogonal to the $v$-lines, i.e., that events in such a hypersurface happen simultaneously for the observers on $v$-lines.  Integration of (\ref{42}) results in 
\begin{equation}
\label{44}
c \, t = c \, T - r_0 \big( F(R)-F(R_p) \big)  
\end{equation}
with
\begin{equation}
\label{45}
F(R) = \dfrac{1}{2} \, \mathrm{log} 
\dfrac{\Big(\sqrt{1-8\mu}+\dfrac{2R}{r_0}-1 \Big)^{(1-8\mu)^{-1/2}-1}
}{
\Big(\sqrt{1-8\mu}-\dfrac{2R}{r_0}+1 \Big)^{(1-8\mu)^{-1/2}+1}
} \, .
\end{equation}
In (\ref{44}) we have chosen the integration constant such that $t$ coincides with $T$ on the photon sphere.

The hypersurfaces $t= \mathrm{const.}$ are shown as solid curves in Fig. \ref{fig:vt}. The time function $t$ runs from $- \infty$ to $+ \infty$ on the domain $\frac{r_0}{2} \big(1- \sqrt{1-8 \mu} \big) < R < \frac{r_0}{2} \big(1+ \sqrt{1-8 \mu} \big)$. Outside of this domain there is a region where $v$ is time-like but $t$ is running backwards; this region is shown in light gray in Fig. \ref{fig:vt}. Also note that $t$ does \emph{not} give proper time along the $v$-lines: The vector field $\partial / \partial v$ is synchronizable but not proper-time synchronizable.

It is now easy to calculate the expansion of the photon sphere as observed by the observers on $v$-lines. If we restrict (\ref{44}) to the hypersurface $R=R_p$ we find
\begin{equation}
\label{46}
    c\, t = c \, T = r_0 \, \mathrm{log} \dfrac{r}{R_p} 
\end{equation}
where the second equality follows from (\ref{4}). Solving for $r$ results in
\begin{equation}
\label{47}
    r = R_p \, e^{ct/r_0} 
\end{equation}
which demonstrates that the observers on $v$-lines see the photon sphere exponentially expanding with time $t$. As $r$ is an area coordinate, this result is in agreement with (\ref{24}).

\begin{figure*}[ht!]
\includegraphics[width = .75\textwidth]{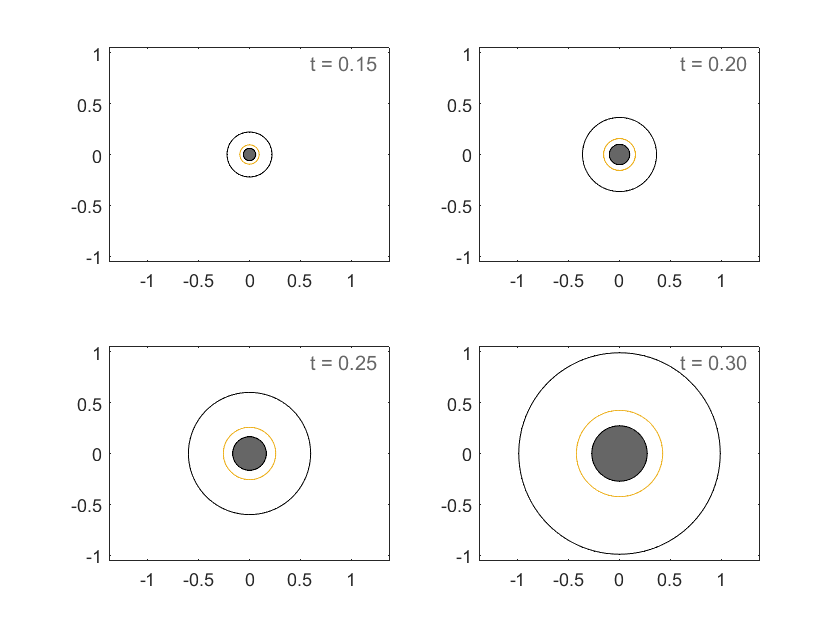} 
\caption{Evolution of photon sphere as seen by observer on a $ v $-line ($ \mu = 1/24 $) }
\label{fig4}
\end{figure*}

Similarly, we can calculate the expansion of any other sphere $R = \mathrm{const.}$ in the domain (\ref{40}). We can do this in particular for the horizons at $R_{\pm}$ which results in
\begin{equation}
\label{48}
    r = R_{\pm} \, e^{F(R_{\pm})-F(R_p)} \, e^{ct/r_0} \, .
\end{equation}
So the horizons expand exponentially, as any other sphere $R= \mathrm{const.}$ does.
We have plotted the expansion of the photon sphere and of the horizons in Fig. \ref{fig4}, where the inner dark region is bounded by the inner horizon, the inner (orange) circle represents the photon sphere and the outer (black) circle represents the outer horizon. Lengths are given in units of $ r_0  $ and times are given in units of $r_0/c$. In our theoretical model the initial radius of the photon sphere is zero at $ t = -\infty $ and its radius becomes infinite at $ t = \infty $.

\begin{figure*}[ht!]
\includegraphics[width = .45\textwidth]{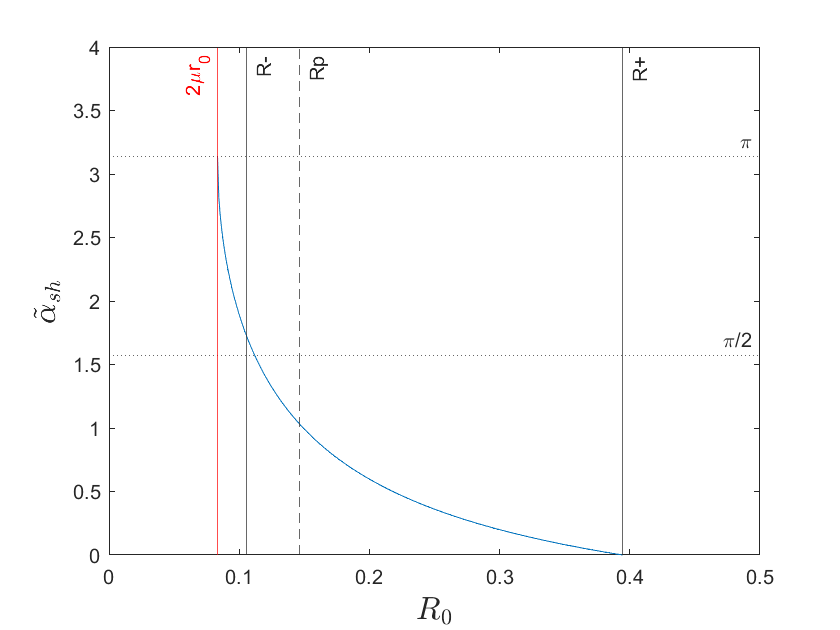} 
\caption{Angular radius of shadow as seen by observer on $ v $-line ($ \mu = 1/24 $) }
\label{tilde alpha}
\end{figure*}

We will now calculate the angular radius of the shadow as seen by an observer on a $v$-line. We call this angular radius $\tilde{\alpha}{}_{\mathrm{sh}}$, and we relate it to the angular radius of the shadow as seen by a conformally static observer, $\alpha _{\mathrm{sh}}$, by the special-relativistic aberration formula 
\begin{equation}
\mathrm{tan} ^2 \Big( \frac{\tilde{\alpha}{}_{\mathrm{sh}}}{2} \Big) = \dfrac{(c-V)}{(c+V)} \, \mathrm{tan} ^2 \Big( \frac{\alpha _{\mathrm{sh}}}{2} \Big)  = \dfrac{(c-V)}{(c+V)} \, \dfrac{\big( 1 - \mathrm{cos} \, \alpha _{\mathrm{sh}} \big)^2}{\mathrm{sin} ^2 \alpha _{\mathrm{sh}}} \, .
\label{eq:aberration}
\end{equation}
Here $\alpha _{\mathrm{sh}}$ is given by (\ref{33}) or (\ref{eq:shadowstat}), and $V$ is the momentary 3-velocity of the observer on a $v$-line with respect to the observer on a $T$-line. The latter has 4-velocity
\begin{equation}
U = \dfrac{e^{-cT/r_0} }{\sqrt{1- \frac{2 \mu r_0}{R}- \frac{2R}{r_0}}} \, \frac{\partial}{\partial T}\, , \quad g(U,U) = - c^2
\end{equation}
whereas, by (\ref{43}), the former has 4-velocity
\begin{equation}
\tilde{U} = \dfrac{e^{-cT/r_0} }{\sqrt{1- \frac{2 \mu r_0}{R}}} \Big( \frac{\partial}{\partial T} - \frac{cR}{r_0} \, \frac{\partial}{\partial R} \Big) \, , \quad g \big( \tilde{U}, \tilde{U} \big) = - c^2 \, .
\end{equation}
By applying, at each moment, the special-relativistic formula
\begin{equation}
g\big( U , \tilde{U} \big) = \dfrac{-c^2}{\sqrt{1-\Big( \frac{V}{c} \Big) ^2 }}    
\label{eq:sr}
\end{equation}
we find that the relative 3-velocity $V$ of the two observers is given by
\begin{equation}
\dfrac{V}{c} = \dfrac{R_0^2}{R_0^2+2 (R_+ - R_0)(R_0 - R_-)}
\label{eq:V}
\end{equation}
at $R=R_0$. Plugging (\ref{eq:V}) and (\ref{33}) into the aberration formula (\ref{eq:aberration}) results in
\begin{equation}
\mathrm{tan} ^2 \Big( \frac{\tilde{\alpha}{}_{\mathrm{sh}}}{2} \Big) = \dfrac{R_0^3 (4 \mu r_0 - R_p)}{R_p^3 (R_0-2\mu r_0 )} \, \big( 1 - \mathrm{cos} \, \alpha _{\mathrm{sh}} \big)^2 \, .
\label{eq:talpha}
\end{equation}
This formula is valid for $R_-<R_0<R_+$., i.e., on the domain where both the $T$-lines and the $v$-lines are time-like. Note that on this domain $R_0>2 \mu r_0$ and that, because of our assumption $0 < \mu < 1/16$, also $4 \mu r_o > R_p$. We read from (\ref{eq:talpha}) that for $R_0 \to R_+$, where $\alpha _{\mathrm{sh}} \to 0$, also $\tilde{\alpha}{}_{\mathrm{sh}} \to 0$. For $R_0 \to R_-$, where $\alpha _{\mathrm{sh}} \to \pi$, however, $\tilde{\alpha}{}_{\mathrm{sh}}$ approaches a finite value. By inserting (\ref{eq:shadowstat}) we see that (\ref{eq:talpha}) can be analytically extended to all values $2 \mu r_0 < R_0 < R_+$, i.e., to the entire domain where the $v$-lines are time-like,
\begin{equation}
\mathrm{tan} ^2 \Big( \frac{\tilde{\alpha}{}_{\mathrm{sh}}}{2} \Big) = \dfrac{R_0^3 (4 \mu r_0 - R_p)}{R_p^3 (R_0-2\mu r_0 )} \, \Bigg( 1 - \dfrac{(R_0-R_p)}{R_0} \sqrt{1 + \dfrac{2 \mu r_0 R_p}{R_0 \big( 4 \mu r_0 - R_p \big)}} \; \Bigg)^2 \, . 
\end{equation}
In Fig. \ref{tilde alpha} $\tilde{\alpha}{}_{\mathrm{sh}}$ is plotted against $R_0$, with the latter given in units of $r_0$. So an observer on a $v$-line sees the shadow come into existence when crossing the outer horizon. The shadow gradually grows while the observer moves inwards. It covers less than half of the sky at $R_0=R_p$, which reflects the fact that aberration has a demagnifying effect in the forward direction. Nothing peculiar happens when the observer crosses the inner horizon. The shadow covers the entire sky in the limit $R \to 2 \mu r_0$, i.e., when the observer reaches the speed of light. 

\section{Photon sphere and shadow of a black hole with decreasing mass}
\label{sec3}
We will now consider the case that the mass function $m(v)$ is decreasing where we will again assume that $m(v)$ is a linear function, $m(v) = - a v$ with a positive constant $a$. A Vaidya metric with a decreasing mass function may be viewed as a simplified model for a black hole that evaporates by emitting Hawking radiation, as was suggested already in the 1980s, see Hiscock \cite{Hiscock1981a,Hiscock1981b}, Kuroda \cite{Kuroda1984a} and Beciu \cite{Beciu1984}. There are two different ways in which an evaporating black hole can be modeled by a Vaidya metric. One possibility is to use (\ref{1}) and (\ref{eq:Tgen}) with the lower sign, i.e., to use outgoing Eddington-Finkelstein-like coordinates. Then the central object would lose mass by emitting a null dust of positive energy density. For a linearly decreasing mass function the space-time region covered by the $(v,r,\theta,\phi)$ coordinates would be just the time-reversed version of the space-time region considered in the previous section. That is to say, all results would literally carry over if we just replace $v$ by $-v$ and $T$ by $-T$ everywhere. In particular, we would again have two horizons in the space-time region covered by our coordinates, with the inner one now being a \emph{white-hole} horizon.

Here we will consider another possibility. We will use (\ref{1}) and (\ref{eq:Tgen}) again with the upper sign. Doing so with a \emph{decreasing} function $m(v)$ means that the null dust is again ingoing but with a negative energy density. We will view such an ingoing null dust with negative energy density as a model for Hawking radiation. Of course, this model is (over-)simplified, in particular because Hawking radiation consists of a superposition of ingoing negative energy density and outgoing positive energy density. Unfortunately, such a superposition cannot be modeled by a single Vaidya metric. (The simplest model that describes ingoing negative energy density and outgoing positive energy density requires patching together two Vaidya metrics, see Hiscock \cite{Hiscock1981b}.)   
In any case, we believe that studying the case of a black hole that loses mass by absorbing negative energy density is of some conceptual interest. We will see that this situation is considerably different from the above-mentioned case which is just the time-reversed version of the accreting black hole treated in the previous section. In particular, we will see that now we have only one horizon instead of two.

According to this plan, we consider the metric 
\begin{equation}
\label{49}
    ds^2 = -\left( 1 + \frac{2 a v }{r} \right)dv^2 + 2 \, dv \, dr + r^2(d\theta^2 + \mathrm{sin}^2\theta \, d\phi^2) 
\end{equation}
where the corresponding energy-momentum tensor (\ref{eq:Tgen}) is
\begin{equation}
    T^{\mu \nu} = - \dfrac{8 \pi G}{c^4} \, a \, K^{\mu} K^{\nu} \, , \qquad
    K^{\mu} \partial _{\mu} = - \partial _r \, .
\end{equation}

The metric is regular for $ -\infty < v < \infty $ and $ 0 < r < \infty $, whereas $ \theta $ and $ \phi $ have their usual range on the sphere as before. However, as the mass is negative for $0<v< \infty$, we consider the metric only on the domain $-\infty <  v < 0$. That is to say, the central object starts with an infinite mass (which is admittedly an idealization) and then radiates until the mass has gone down to zero. From this moment on we think of the space-time as being of some other form which we need not specify because it is irrelevant for the shadow. 

We have already emphasized that, in our view, the calculations in this Section are of some conceptual interest. In view of astrophysical observations, the effect is so small that there is no chance to observe it with stellar or supermassive black holes: It is known that for a black hole of one Solar mass Hawking radiation produces a mass loss of $dM/d \tau \approx - 10^{-67} M_{\odot}/\mathrm{yr}$. As the mass loss scales with $M^{-2}$, it is even 20 orders of magnitude smaller for a supermassive black hole of $10^{10}$ Solar masses.  

\subsection{Light-like geodesics in the equatorial plane}
\label{subsec3A}
As in the case of increasing mass, the metric admits a conformal Killing vector field, i.e., we can make a cordinate transformation $(v, r, \theta, \phi)$ $\rightarrow$ $ (T, R, \theta, \phi) $ such that the metric becomes manifestly conformally static. In the case at hand, this transformation reads
\begin{equation}
\label{50}
     v = - r_0 \, e^{-cT/r_0} \quad\mathrm{and}\quad r = R \, e^{-cT/r_0}
\end{equation}
where $ r_0 $ is a constant and $ c $ is the speed of light. In the new coordinates the metric (\ref{49}) takes the following form:
\begin{equation}
\label{51}
     ds^2 = e^{-2cT/r_0}\left\{-\left( 1 - \frac{2 a r_0 }{R} + \frac{2R}{r_0} \right)c^2 dT^2 + 2 \, c \, dT \, dR + R^2(d\theta^2 + {\mathrm{sin}}^2\theta \, d\phi^2)\right\} \, .
\end{equation}
The metric (\ref{51}) is regular on the domain $ -\infty < T < \infty $ and $ 0 < R < \infty $.  From equation (\ref{50}) it can be seen that on this domain the original coordinates have the range of $ -\infty < v < 0$ and $ 0 < r < \infty $. This is precisely the domain to which we restrict our consideration.

The conformal Killing vector field  $ \partial / \partial T $ is time-like if

\begin{equation}
    \label{52}
     1 - \frac{2 a r_0 }{R} + \frac{2R}{r_0} > 0
\end{equation}
i.e., on the domain
\begin{equation}
    \label{53}
     R > R_h
\end{equation}
where
\begin{equation}
    \label{54}
    R_h =  \frac{r_0}{4}(-1 + \sqrt{1 + 16 a})
\end{equation}
is the position of the horizon. Again, it is obvious
that this is a conformal Killing horizon. Thus, in this case there is only one horizon, for any $a > 0$, instead of two as in the previous case. 

\begin{figure*}[ht!]
\includegraphics[width = .65\textwidth]{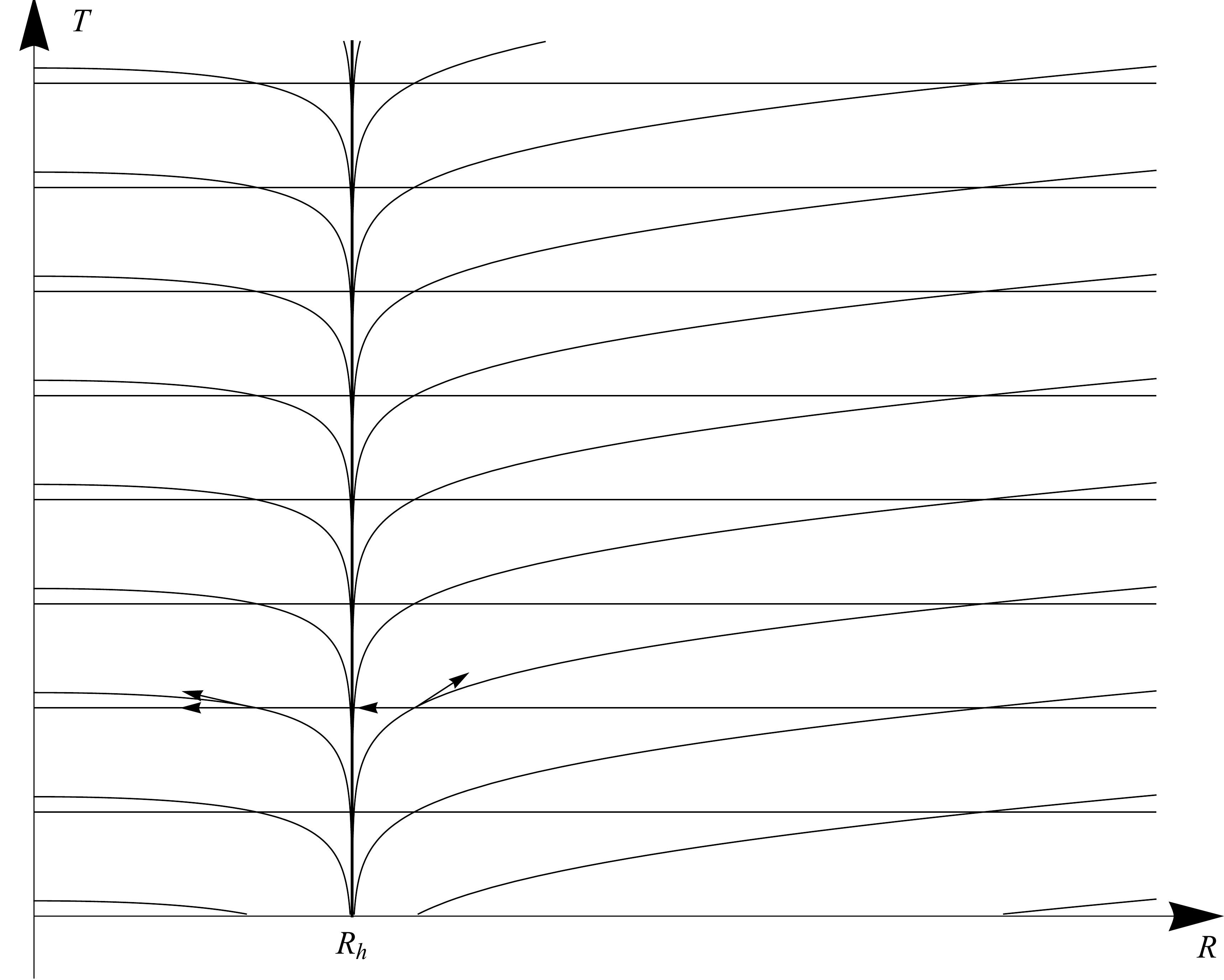}
\caption{{Radial light rays}}
\label{fig:radial2}
\end{figure*}

The light-like geodesics in the equatorial plane obey the following condition derived from (\ref{51}):
\begin{equation}
\label{55}
   0 =  -\left( 1 - \frac{2 a r_0 }{R} + \frac{2R}{r_0} \right)c^2 \dot{T}^2 + 2 c \dot{T}\dot{R} + R^2\dot{\phi}^2 \, .
\end{equation}
The conformal Killing vector field $ \partial / \partial T $ and the Killing vector field $\partial / \partial \phi$ give us the following two constants of motion $E$ and $L$: 
\begin{equation}
\label{56}
    -\left( 1 - \frac{2 a r_0 }{R} + \frac{2R}{r_0} \right) e^{-2cT/r_0} c \dot{T}  + e^{-2cT/r_0} \dot{R} = E \, ,
\end{equation}
\begin{equation}
\label{57}
    L = e^{-2cT/r_0}R^2 \dot{\phi} \, .
\end{equation}
From (\ref{55}) we read that \emph{radial} light rays ($\dot{\phi}=0$) satisfy \begin{equation}
    c \, dT =0 \, , \quad c \, dT = \frac{2dR}{1 - \frac{2 a r_0 }{R} + \frac{2R}{r_0}}
\label{eq:dradial2}
\end{equation}
where the first equation is for ingoing and the second for outgoing radial light rays. Integration of these equations results in
\begin{equation}
    c \, T = \mathrm{const.} \, , \quad c \, T = \frac{r_0}{k + R_h}\left\{ k \,  \mathrm{log} \left(\frac{R}{k}+1\right) +  R_h \mathrm{log} \left(\frac{R}{R_h} - 1\right) \right\} + \mathrm{const.}
\label{eq:radial2}
\end{equation}
with $ R_h $ from (\ref{54}) and 
\begin{equation}
k =\frac{r_0}{4} \left( 1 + \sqrt{1 + 16a} \right) \, .
\label{eq:k}
\end{equation}
The ingoing and outgoing radial light-like geodesics are plotted in Fig. \ref{fig:radial2}. Again, the little arrows indicate the future direction.

For non-radial light rays we divide equations (\ref{55}) and (\ref{56}) by $ \dot{\phi} ^2 $ and $\dot{\phi}$, respectively, and use equation (\ref{57}) to find
\begin{equation}
\label{58}
    -\left( 1 - \frac{2 a r_0 }{R} + \frac{2R}{r_0} \right)c^2 \left(\frac{dT}{d\phi}\right)^2 + 2 c \frac{dT}{d\phi} \frac{dR}{d\phi} + R^2 = 0 \, ,
\end{equation}

\begin{equation}
\label{59}
    -\left(1 - \frac{2 a r_0 }{R} + \frac{2R}{r_0} \right)c \frac{dT}{d\phi} + \frac{dR}{d\phi} = \frac{E R^2}{L} \, .
\end{equation}
By solving equations (\ref{58}) and (\ref{59}) for $ \frac{dT}{d\phi} $ and $ \frac{dR}{d\phi} $, we obtain the following equations describing light-like geodesics in the equatorial plane for the metric (\ref{51}):

\begin{equation}
\label{60}
    \frac{dR}{d\phi} = \pm \sqrt{\frac{E^2 R^4}{L^2} - R^2 + 2a r_0 R - \frac{2R^3}{r_0}} \, ,
\end{equation}

\begin{equation}
\label{61}
    c\frac{dT}{d\phi} = \frac{-\frac{E R^2}{L} \pm \sqrt{\frac{E^2 R^4}{L^2} - R^2 + 2a r_0 R - \frac{2R^3}{r_0}}}{1 - \frac{2 a r_0 }{R} + \frac{2R}{r_0}} \, .
\end{equation}
Eq. (\ref{60}) determines $ R $ as a function of $ \phi $. Thereupon, (\ref{61}) can be used to determine $ T $ as a function of $ \phi $.

\subsection{The photon sphere}
\label{subsec3B}
\begin{figure*}[ht!]
\includegraphics[width = .45\textwidth]{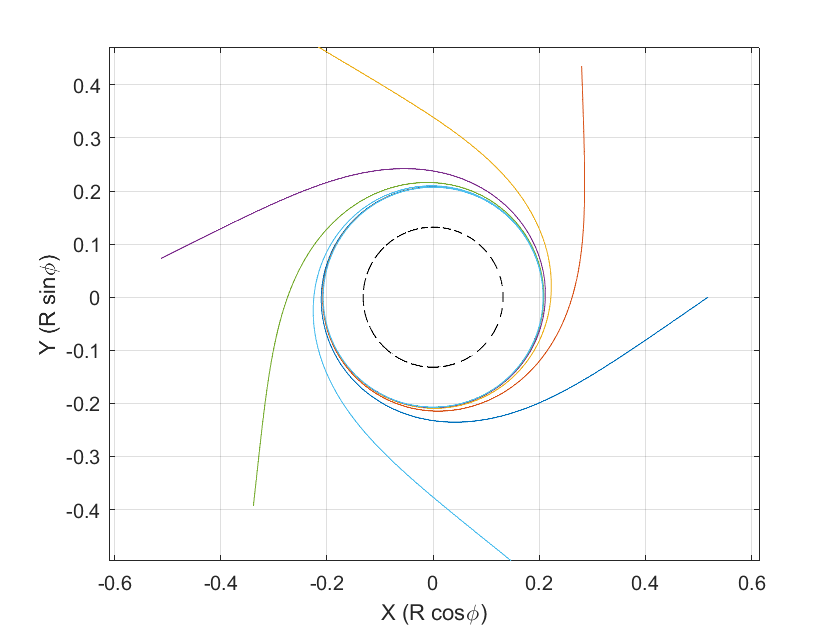}
\caption{Light rays spiraling towards the photon sphere ($ a = 1/12 $)}
\label{fig5}
\end{figure*}

We will now show with the help of (\ref{60}) that the metric (\ref{51}) admits exactly one photon sphere, located outside of the horizon. As mentioned in the previous case, the condition $ \frac{dR}{d\phi} = 0 $ gives the extremum points of light paths, $ R = R_m $. Then the radius coordinate $ R_p $ of the photon sphere can be calculated from the additional condition $ \frac{d^2R}{d\phi^2} = 0 $. By (\ref{60}) the condition $ \frac{dR}{d\phi} = 0 $ holds at $ R = R_m $ if
\begin{equation}
\label{62}
    \frac{E^2 R_m^3}{L^2} - R_m + 2a r_0 -  \frac{2R_m^2}{r_0} = 0 \, .
\end{equation}
This equation allows us to express $ \frac{E^2}{L^2} $ in terms of $R_m$,
\begin{equation}
\label{63}
    \frac{E^2}{L^2} = \frac{R_m - 2a r_0 + \frac{2 R_m^2}{r_0}}{R_m^3} \, .
\end{equation}
Now we compute $\frac{d^2 R}{d\phi^2}$ at $ R = R_m $ from (\ref{60}):
\begin{equation}
\label{64}
    \frac{d^2R}{d\phi^2}\bigg|_{R=R_m} = \frac{4E^2R_m^3}{L^2} - 2R_m + 2 a r_0 - \frac{6R_m^2}{r_0} \, .
\end{equation}
From (\ref{63}), equation (\ref{64}) can be written as
\begin{equation}
\label{65}
    \frac{d^2R}{d\phi^2}\bigg|_{R=R_m} = \frac{2}{r_0}\left\{ R_m - \frac{r_0}{2}(-1 - \sqrt{1 + 12a})  \right\}\left\{ R_m - \frac{r_0}{2} (-1 + \sqrt{1 + 12a})   \right\} \, .
\end{equation}
To find the location of the photon sphere, we have to equate the right-hand side of (\ref{65}) to zero. As $ R_m  > 0 $, the only solution $ R_m = R_p $ is

\begin{equation}
\label{66}
    R_p = \frac{r_0}{2} (-1 + \sqrt{1 + 12a}) \, .
\end{equation}
This is the only possible value for a photon sphere. We still have to check if the condition $\frac{E^2}{L^2} > 0$ is satisfied at $R_m=R_p$. By rewriting equation (\ref{63}) as 

\begin{equation}
\label{67}
    \frac{E^2}{L^2} = \frac{2}{r_0}\frac{\left(R_m - \frac{r_0}{4}(-1 - \sqrt{1 + 16a}) \right)\left( R_m - \frac{r_0}{4}(-1 + \sqrt{1 + 16a}) \right)}{R_m^3} \, ,
\end{equation}
we see that this condition requires $ R_m > \frac{r_0}{4}(-1 + \sqrt{1 + 16a}) $ which is indeed satisfied by $ R_m = R_p$ . We summarize these findings in the following way: For any choice of the parameter $a>0$, there is one horizon, at radius $ R_h  $ given by (\ref{54}), and one photon sphere, at radius $R_p$ given by (\ref{66}) which is always between the horizon and infinity, see Fig. \ref{fig5}. 

To find the relation between $T$ and $\phi$ for an equatorial light-like geodesics in the photon sphere we insert $ R_m = R_p $ into (\ref{63}) which yields

\begin{equation}
    \label{68}
    \frac{E}{L} = \frac{\pm \sqrt{1 - \frac{2 a r_0}{R_p} + 2\frac{R_p}{r_0}}}{R_p} \, .
\end{equation}
By (\ref{61}) gives us the desired relation,   

\begin{equation}
\label{69}
c T = \frac{\pm \, R_p \, \phi}{\sqrt{1 - \frac{2 a r_0 }{R_p} + \frac{2R_p}{r_0}}} \, . 
\end{equation}
As $T$ runs over all of $\mathbb{R}$, so does $\phi$. By (\ref{57}), this implies that also in this case the light-like geodesics in the photon sphere are non-extendable: The affine parameter starts at $- \infty$ and then runs up to a value where the singularity is reached. We repeat that this observation is non-trivial because the coordinates $(T,R,\phi,\theta)$ cover only a part of the original space-time (\ref{49}).

Equation (\ref{68}) gives us the ratio of constants $ \frac{E}{L} $ for equatorial light-like geodesics that are completely contained in the photon sphere, and also of such geodesics that spiral asymptotically towards the photon sphere, see Fig. \ref{fig5}. In this figure the dashed line represents the horizon and lengths are given in units of $r_0$. As in the previous case, the $R$-coordinate of the photon sphere is fixed (as is the $R$-coordinate of the horizon), but by (\ref{51}) the area $A$ of the photon sphere depends on $T$,
\begin{equation}
\label{70}
    A=4 \pi e^{-2cT/r_0} R_p^2 = \pi e^{-2cT/r_0} r_0^2 \big(-1 + \sqrt{1 + 12 a} \, \big)^2 \, ,
\end{equation}
so it goes to $ \infty $ for $T \to - \infty$ and to $ 0 $ for $T \to \infty$.

\subsection{The angular radius of the black-hole shadow}
\label{subsec3C}
In this Section we calculate the angular radius of the black-hole shadow as it is seen by an observer on a $ T $-line. Of course, the observer must be between the horizon and infinity. We will see that the angular radius is time-independent, although the area of the photon sphere decreases with $ T $. We use the following tetrad field on the domain $  R > R_h.  $
\begin{equation}
\label{71}
    e_0 = \frac{e^{cT/r_0}}{\left(\sqrt{ 1 - \frac{2 a r_0 }{R} + \frac{2R}{r_0}}\right)c}\frac{\partial}{\partial T} \, ,
\end{equation}

\begin{equation}
\label{72}
    e_1 = \frac{e^{cT/r_0}}{\left(\sqrt{ 1 - \frac{2 a r_0 }{R} + \frac{2R}{r_0}}\right)c}\frac{\partial}{\partial T} + \left(\sqrt{ 1 - \frac{2 a r_0 }{R} + \frac{2R}{r_0}}\right)e^{cT/r_0}\frac{\partial}{\partial R} \, ,
\end{equation}

\begin{equation}
\label{73}
    e_2 = \frac{e^{cT/r_0}}{R}\frac{\partial}{\partial \theta} \quad\mathrm{and}\quad e_3 = \frac{e^{cT/r_0}}{R sin\theta}\frac{\partial}{\partial \phi} \, .
\end{equation}

\begin{figure*}[ht!]
\includegraphics[width = .5\textwidth]{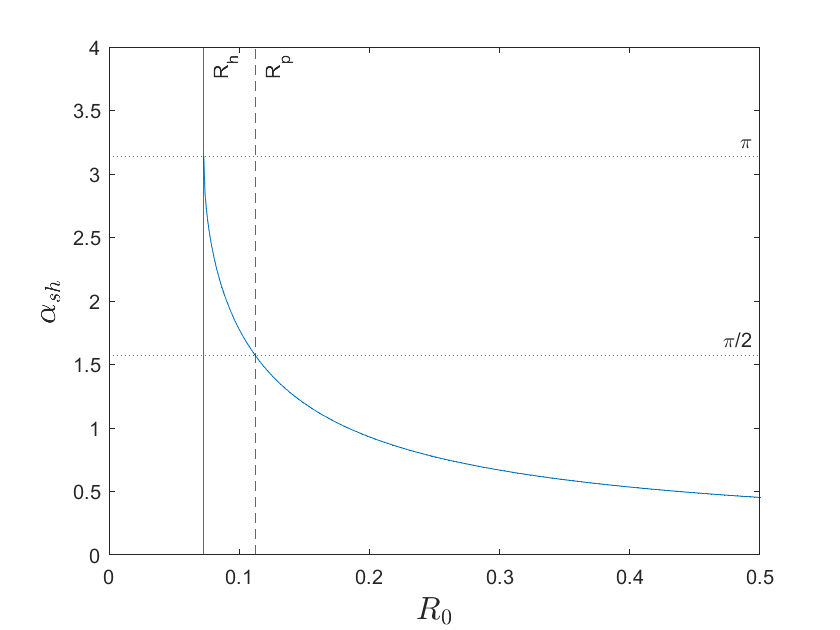}
\caption{Angular radius of the shadow ($ a = 1/24 $)}
\label{fig6}
\end{figure*}

Expanding the tangent vector of an equatorial light-like geodesic $(T(\lambda),R(\lambda),\phi(\lambda))$,  where $ \lambda $ is an affine parameter, with respect to this tetrad results in
\begin{equation}
\label{74}
    \dot{T}\frac{\partial}{\partial T} + \dot{R}\frac{\partial}{\partial R} + \dot{\phi}\frac{\partial}{\partial \phi} = \xi \left( e_0 + e_1 \mathrm{cos} \, \alpha - e_3  \mathrm{sin} \, \alpha \right)
\end{equation}
where $ \xi $ is a scalar factor and $ \alpha $ is the angle between the light-like geodesic and the radial direction in the rest system of the observers with four-velocity $e_0$. Comparing the coefficients of $\frac{\partial}{\partial R}$ and $\frac{\partial}{\partial \phi}$ in (\ref{74}) yields
\begin{equation}
\label{75}
    \dot{R} = \mathrm{cos} \, \alpha \left(\sqrt{ 1 - \frac{2 a r_0 }{R} + \frac{2R}{r_0}}\right)e^{cT/r_0} \, ,
\end{equation}
\begin{equation}
\label{76}
    \dot{\phi} = -\frac{(\mathrm{sin} \, \alpha) e^{cT/r_0}}{R} \, .
\end{equation}
From equations (\ref{60}), (\ref{75}) and (\ref{76}), we have
\begin{equation}
\label{77}
    \frac{1}{\frac{E^2 R^4}{L^2} - R^2 + 2a r_0 R - \frac{2R^3}{r_0}} = \frac{\mathrm{sin}^2\alpha}{\mathrm{cos}^2\alpha\left( 1 - \frac{2 a r_0 }{R} + \frac{2R}{r_0}\right)R^2}
\end{equation}
With the help of (\ref{63}) and (\ref{77}), we  write $\mathrm{sin} \, \alpha $ as follows:
\begin{equation}
\label{78}
    \mathrm{sin} \, \alpha = \sqrt{\frac{R_m^3\left(R - 2a r_0 + \frac{2R^2}{r_0}\right)}{R^3\left(R_m - 2a r_0 + \frac{2R_m^2}{r_0}\right)}} \, .
\end{equation}
The angular radius $\alpha _{\mathrm{sh}}$ of the shadow for an observer at $ R = R_0 $ is found by inserting into this equation the radius $R_m=R_p$ of the photon sphere,
\begin{equation}
\label{79}
    \mathrm{sin} \, \alpha_{\mathrm{sh}} = \sqrt{\frac{R_p^3\left(R_0 - 2a r_0 + \frac{2R_0^2}{r_0}\right)}{R_0^3\left(R_p - 2a r_0 + \frac{2 R_p^2}{r_0}\right)}} = \sqrt{\frac{R_p^3 (R_0 + k) (R_0-R_h)}{R_0^3 (R_p+k) (R_p-R_h)}}
\end{equation}
with $R_h$ from (\ref{54}) and $ k $ from (\ref{eq:k}). This equation is valid for all observer positions $ R_0 > R_h$ where the expression under the square-root is indeed between 0 and 1. Eq. (\ref{79}) must be supplemented with the information that $\alpha _{\mathrm{sh}}$ is in the interval between 0 and $\pi/2$ for $ R_0>R_p$ and in the interval between $\pi/2$ and $\pi$ for $R_h<R_0<R_p$. We have $\alpha_{\mathrm{sh}} \to 0$ (bright sky) for $R_0 \to \infty$ and $\alpha _{\mathrm{sh}} \to \pi$ (dark sky) for $R_0 \to R_h$. At $R_0=R_p$ the shadow covers half of the sky which is obvious without any calculation for this case too. Fig. \ref{fig6} shows the angular radius of the black-hole shadow $ \alpha_{\mathrm{sh}} $ as a function of $ R_0 $. The solid black line labeled $ R_h $  marks the horizon, the dashed black line labeled $ R_p $ represents the photon sphere and $R_0$ is given in units of $r_0$. For this case the domain of interest is outside of the horizon because inside the horizon no conformally static observer can exist.  

As for the accreting black hole, is it also possible to give a formula for the cosine, rather than the sine, of $\alpha _{\mathrm{sh}}$,
\begin{equation}
    \mathrm{cos} \, \alpha_{sh} = \dfrac{(R_0-R_p)}{R_0} \sqrt{1+ \frac{2 a r_0 R_p}{R_0 (4 a r_0 - R_p)}} \, .
\label{eq:shadowstat2}
\end{equation}
This equation determines $\alpha _{\mathrm{sh}}$ uniquely in the interval between 0 and $\pi$, for all $R_h<R_0<\infty$.  

In contrast to the case of an accreting black hole, we don't have an outer horizon, so we can consider conformally static observers that are arbitrarily far away from the black hole. For $R_0 \gg R_p$ we may rewrite eq. (\ref{79}) as
\begin{equation}
    \mathrm{sin} \, \alpha_{\mathrm{sh}} = \sqrt{\dfrac{2 \, R_p^2}{ R_p r_0-2 a r_0^2 + 2 R_p^2}} \; \sqrt{\dfrac{R_p}{R_0} } \, \Bigg( 1 + \dfrac{r_0}{4 R_p} \, \dfrac{R_p}{R_0} + O \Big( \dfrac{R_p^2}{R_0^2} \Big) \Bigg) 
\end{equation}
and neglect the terms of second and higher order. We see that the sine of the angular radius of the shadow falls off with $R_0^{-1/2}$ in leading order. This is in contrast to the Schwarzschild case where it falls off, for a static observer at Schwarzschild coordinate $R_0$, with $R_0^{-1}$, see Synge \cite{10.1093/mnras/131.3.463}.

\subsection{The gravitational red-shift of light}
\label{subsec3D}
In this section we compute the gravitational red-shift of light in the metric (\ref{51}). We assume that two observers, labeled $ A $ and $ B $ respectively, are moving on $ T $-lines in the domain outside of the horizon. Then 
\begin{equation}
\label{80}
    \frac{\nu_A}{\nu_B} = \sqrt{\frac{g_{00}(T_B, R_B)}{g_{00}(T_A, R_A)}}
\end{equation}
where $ \nu_A $ and $ \nu_B $ are the frequency of the light at $ (T_A, R_A, \theta_A, \phi_A) $ and $ (T_B, R_B, \theta_B, \phi_B) $, respectively. From (\ref{51}), equation (\ref{80}) becomes
\begin{equation}
\label{81}
     \frac{\nu_A}{\nu_B} = \frac{\sqrt{ 1 - \frac{2 a r_0 }{R_B} + \frac{2R_B}{r_0}}}{\sqrt{ 1 - \frac{2 a r_0 }{R_A} + \frac{2R_A}{r_0}}}e^{c(T_A-T_B)/r_0} \, .
\end{equation}
As in the previous case, the fact that $ \partial / \partial T $ is a conformal Killing vector field implies that $ T_B - T_A $ has the same value for all light rays traveling from $ A $ to $ B $, so the red-shift is again time-independent. 

For light rays traveling in the radial direction, which requires $(\theta_A, \phi_A) = (\theta_B, \phi_B)$, we find from (\ref{eq:radial2}) that (\ref{81}) becomes

\begin{equation}
\label{83}
     \frac{\nu_A}{\nu_B} = \frac{\sqrt{ 1 - \frac{2 a r_0 }{R_B} + \frac{2R_B}{r_0}}}{\sqrt{ 1 - \frac{2 a r_0 }{R_A} + \frac{2R_A}{r_0}}}
\end{equation}
for ingoing and 

\begin{equation}
\label{85}
     \frac{\nu_A}{\nu_B} = \frac{\sqrt{ 1 - \frac{2 a r_0 }{R_B} + \frac{2R_B}{r_0}}}{\sqrt{ 1 - \frac{2 a r_0 }{R_A} + \frac{2R_A}{r_0}}}\left\{ \left( \frac{k + R_A}{k+R_B} \right)^{\frac{k}{k + R_h}} \left( \frac{R_A - R_h}{R_B - R_h} \right)^{\frac{R_h}{k + R_h}}  \right\}
\end{equation}
for outgoing light rays, where $ k =\frac{r_0}{4} \left( 1 + \sqrt{1 + 16a} \right)$ and $ R_h =\frac{r_0}{4} \left(- 1 + \sqrt{1 + 16a} \right)$.

\subsection{Analysis in the original coordinates}
\label{subsec3E}
As in the previous case, we now transform back the results to the original coordinate system $(v, r, \theta, \phi)$. This allows us to analyze the shadow as seen by observers on $v$-lines for the metric (\ref{49}). Recall that we restrict to the domain where $- \infty < v < 0$ (and $0 < r < \infty$). 

From (\ref{49}), we read that the $v$-lines are time-like on the domain
\begin{equation}
\label{86}
  - 2 a v < r \, , \quad 2 a r_0 < R  \, .
\end{equation}
This domain includes the photon sphere if $0 < a < 1/4$; it always lies outside of the horizon. As in the previous case we observe that the hypersurfaces $v = \mathrm{const.}$ are light-like everywhere and that $r$ is an \emph{area coordinate}, i.e., that the sphere $(v,r)= \mathrm{const.}$ has area $4 \pi r^2$, as can be read from (\ref{49}).

\begin{figure*}[ht!]
\includegraphics[width = .45\textwidth]{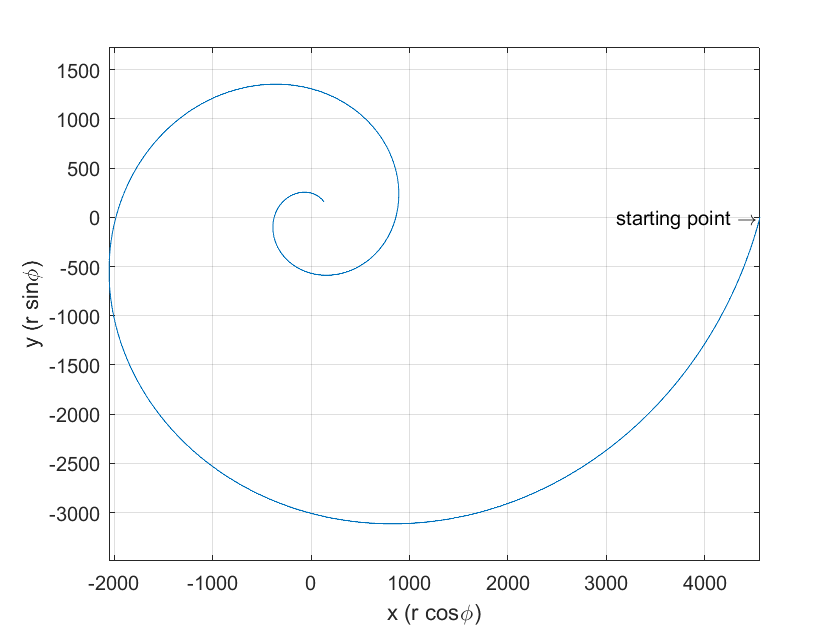} 
\caption{Light path in the photon sphere seen in $(v, r, \theta, \phi)$ coordinates ($ a = 1/24 $)}
\label{fig7}
\end{figure*}

We have found above that in the $ (T, R, \theta, \phi) $ coordinates light paths in the photon sphere are at $R=R_p= \frac{r_0}{2}(-1 + \sqrt{1 + 12 a}) $ and that in the equatorial plane their $T$ and $\phi$ coordinates are related by (\ref{69}). Transforming these two equations back to the original coordinates $(v, r, \theta, \phi)$ with the help of (\ref{50}) results in

\begin{equation}
\label{87}
    r = \frac{-R_p \, v}{r_0} \quad\mathrm{and}\quad \phi =  \pm \left(\frac{r_0}{R_p}\sqrt{1 - \frac{2a r_0}{R_p} + \frac{2R_p}{r_0}}\right) \mathrm{log}\left( \frac{- v}{r_0} \right) \, .
\end{equation}
These  equations  give  us  the  path  of  the  geodesic  parametrized  by $v$. We  see that  in  these  coordinates  the  light-like  geodesics  in  the  photon  sphere  spiral  inwards,  see  Fig. \ref{fig7}, i.e.,  that  the photon  sphere  shrinks. The  radius coordinate $r$, which is again plotted in units of $r_0$, is  linearly  decreasing with $v$.   The  angle $ \phi $ is  a  logarithmic  function  of $ - v $,  so  the angular speed $\frac{d\phi}{d v}$ is inversely proportional to $  - v $ implying that the angular speed increases with time.

We repeat that the hypersurfaces $v = \mathrm{const.}$ are light-like, i.e., that they cannot be interpreted as equal-time slices. For describing the shrinking of the entire photon sphere, as observed by observers on $v$-lines, we have to introduce an appropriate time function for this case also. We define such a function $t$ in the $(T,R,\theta,\phi)$ coordinates by the equation
\begin{equation}
\label{88}
    c \, dt = c \, dT - \dfrac{dR}{1 - \dfrac{2 a r_0}{R} + \dfrac{R}{r_0}} \, .
\end{equation}
With the help of the relation
\begin{equation}
\label{89}
    \dfrac{\partial}{\partial v} = e^{cT/r_0} \dfrac{1}{c} \Big( \dfrac{\partial}{\partial T}+ \dfrac{cR}{r_0} \dfrac{\partial}{\partial R} \Big) \, ,
\end{equation}
which follows from (\ref{50}), it is easy to verify that the hypersurfaces $t = \mathrm{const.}$ are orthogonal to the $v$-lines, i.e., that events in such a hypersurface happen simultaneously for the observers on $v$-lines.  As in the previous case, the vector field $\partial / \partial v$ is synchronizable on the domain (\ref{86}) but not proper-time synchronizable, so $t$ does not give proper time for the observers on $v$-lines.

Integration of (\ref{88}) results in 
\begin{equation}
\label{90}
c \, t =  c \, T - r_0 \big( F(R)-F(R_p) \big)  
\end{equation}
with

\begin{equation}
\label{91}
F(R) = \dfrac{1}{2} \, \mathrm{log} 
\dfrac{\Big(1+\sqrt{1+8a}+\dfrac{2R}{r_0}\Big)^{(1+8a)^{-1/2}+1}
}{
\Big(1-\sqrt{1+8a}+\dfrac{2R}{r_0}\Big)^{(1+8a)^{-1/2}-1}
} \, .
\end{equation}
If we restrict (\ref{90}) to the hypersurface $R=R_p$ we find that
\begin{equation}
\label{92}
    c\, t = c \, T = - r_0 \, \mathrm{log} \dfrac{r}{R_p} 
\end{equation}
where the second equality follows from (\ref{50}). Solving for $r$ results in
\begin{equation}
\label{93}
    r = R_p \, e^{-ct/r_0} 
\end{equation}
which demonstrates that the observers on $v$-lines see the photon sphere exponentially shrinking with time $t$.

\begin{figure*}[ht!]
\includegraphics[width = .75\textwidth]{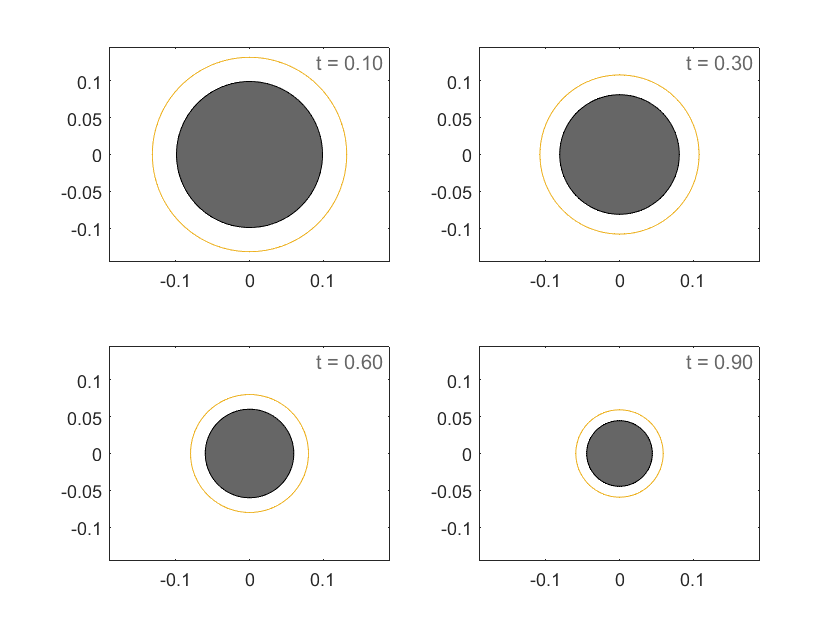} 
\caption{Evolution of photon sphere as seen by observer on $ v $-line ($ a = 1/18 $)}
\label{fig8}
\end{figure*}

Any other sphere $R = \mathrm{const.}$ in the domain (\ref{86}) shrinks in a similar way. In particular, for the horizon we find that 
\begin{equation}
\label{94}
    r = R_{h} \, e^{F(R_{p})-F(R_h)} \, e^{-ct/r_0} \, .
\end{equation}
We have plotted the shrinking of the photon sphere and of the horizon in Fig. \ref{fig8}. The dark region is bounded by the horizon and the (orange) circle represents the photon sphere. Lengths are given in units of $r_0$ and times are given in units of $r_0/c$.

\begin{figure*}[ht!]
\includegraphics[width = .275\textwidth]{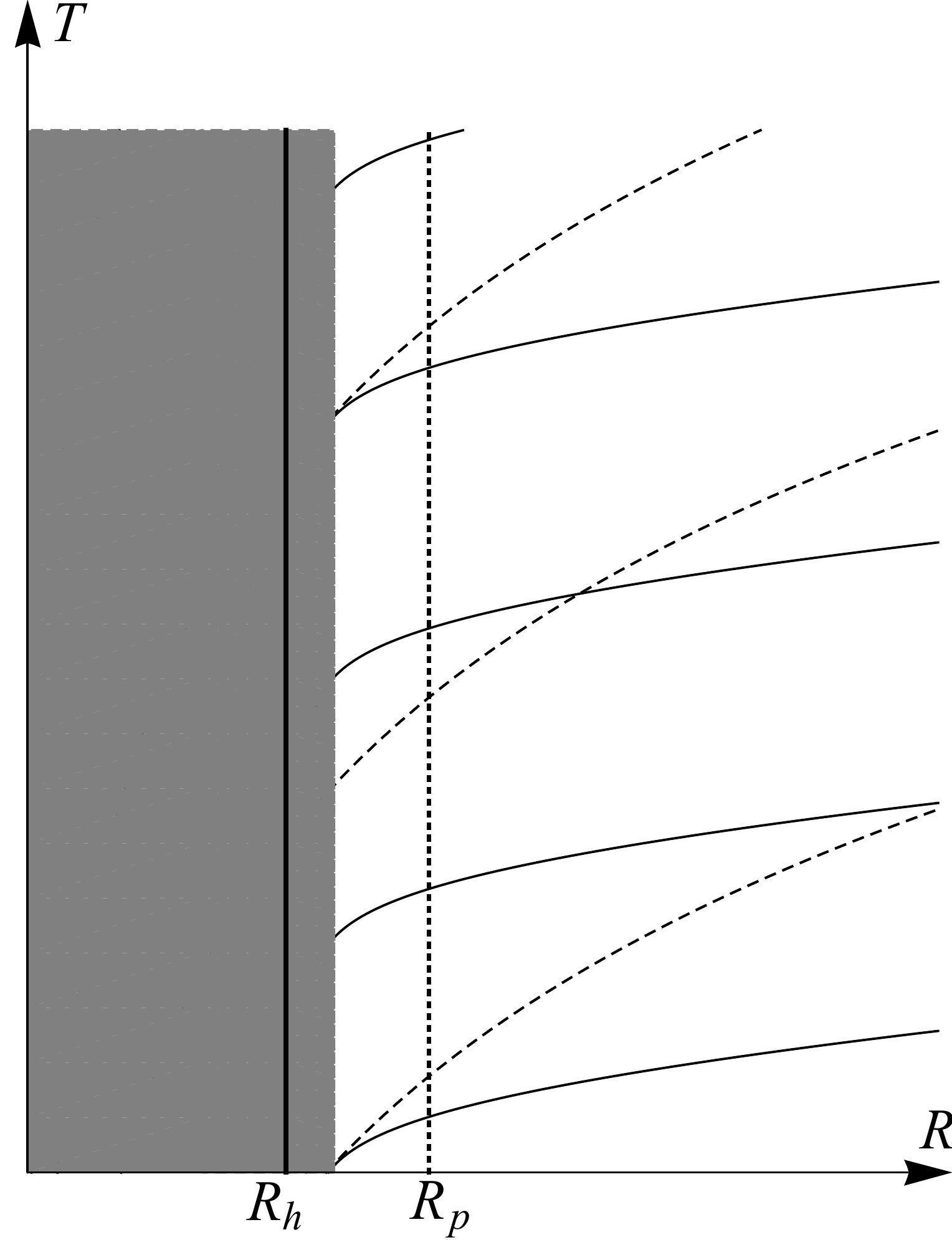} 
\hspace{0.06\textwidth}
\includegraphics[width = .375\textwidth]{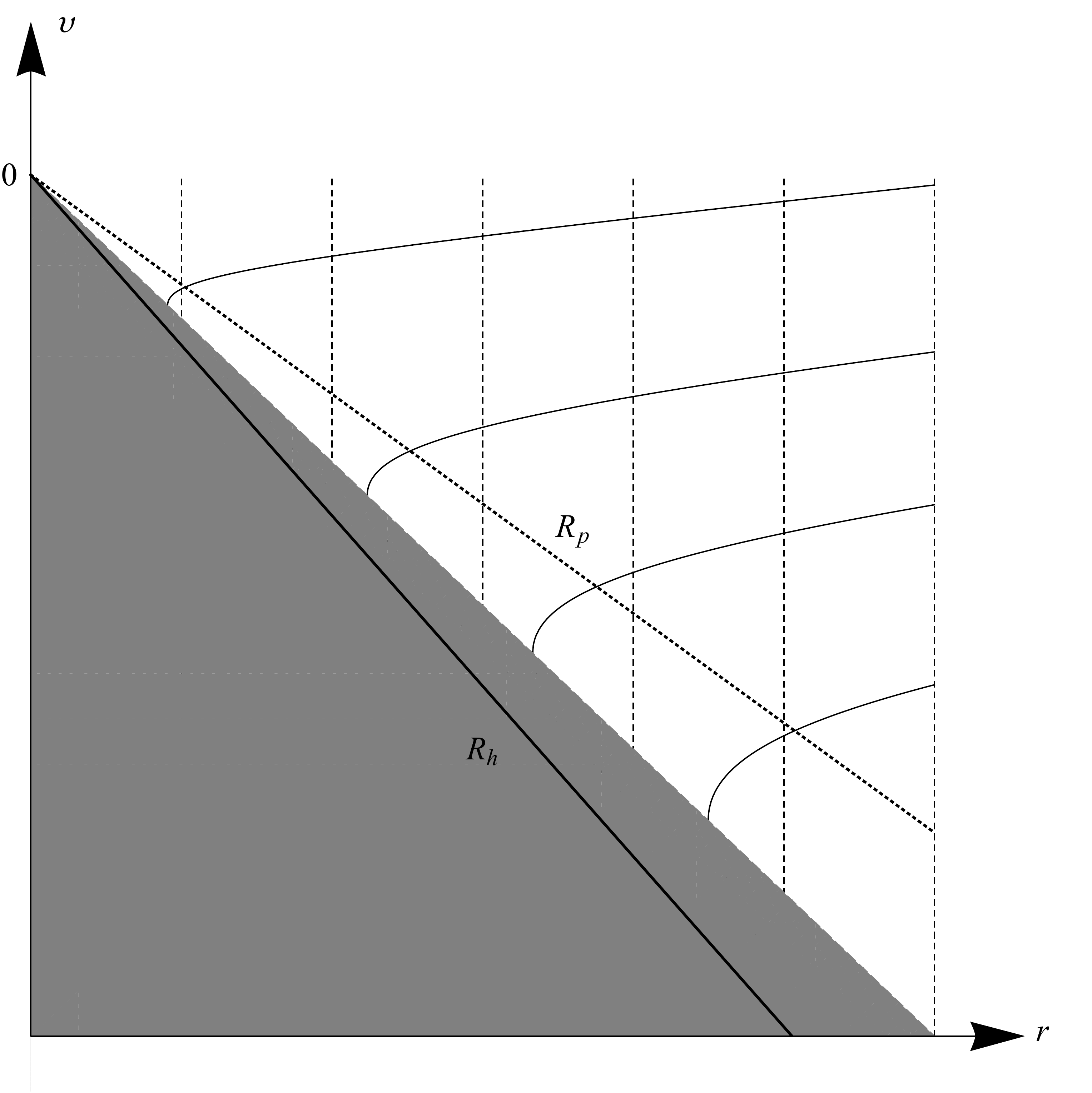} 
\caption{$v$-lines (dashed) and hypersurfaces $t = \mathrm{const.}$ (solid)}
\label{fig:vto}
\end{figure*}

Fig. \ref{fig:vto} shows the $v$-lines and the hypersurfaces $t = \mathrm{const.}$ for the case $a=1/18$. The region where the $v$-lines fail to be time-like is shaded. For the chosen value of $a$ the photon sphere is outside of this region.

We will now calculate the angular radius  $\tilde{\alpha}{}_{\mathrm{sh}}$ of the shadow as seen by an observer on a $v$-line. It is related to the angular radius $\alpha _{\mathrm{sh}}$ of the shadow as seen by a conformally static observer by the special-relativistic aberration formula (\ref{eq:aberration}). Here for this case $\alpha _{\mathrm{sh}}$ is given by (\ref{79}) or (\ref{eq:shadowstat2}), and $V$ is the momentary 3-velocity of the observer on a $v$-line with respect to the observer on a $T$-line. The latter has 4-velocity
\begin{equation}
U = \dfrac{e^{cT/r_0} }{\sqrt{1- \frac{2 a r_0}{R} + \frac{2R}{r_0}}} \, \frac{\partial}{\partial T}\, , \quad g(U,U) = - c^2
\label{eq:U2}
\end{equation}
whereas, by (\ref{89}), the former has 4-velocity
\begin{equation}
\tilde{U} = \dfrac{e^{cT/r_0} }{\sqrt{1- \frac{2 a r_0}{R}}} \Big( \frac{\partial}{\partial T} + \frac{cR}{r_0} \, \frac{\partial}{\partial R} \Big) \, , \quad g \big( \tilde{U}, \tilde{U} \big) = - c^2
\label{eq:Utilde2}
\end{equation}
From (\ref{eq:sr}) we find that
\begin{equation}
\dfrac{V}{c} = \dfrac{-R_0^2}{R_0^2-2 a r_0^2+r_0R_0}
\label{eq:V2}
\end{equation}
at $R=R_0$. Note that here $V$ is negative because, with respect to $U$, the four-velocity $\tilde{U}$ is directed away from the center. 

Plugging (\ref{eq:V2}) and (\ref{79}) into the aberration formula (\ref{eq:aberration}) results in
\begin{equation}
\mathrm{tan} ^2 \Big( \frac{\tilde{\alpha}{}_{\mathrm{sh}}}{2} \Big) = \dfrac{ R_0^3 (4 a r_0 - R_p)}{ R_p^3 (R_0 - 2 a r_0)} \, \big( 1 - \mathrm{cos} \, \alpha _{\mathrm{sh}} \big)^2 
\label{eq:talpha2}
\end{equation}
This formula is valid for $R_0>2 a r_0$., i.e., on the domain where both the $T$-lines and the $v$-lines are time-like. We read from (\ref{eq:talpha2}) that for $R_0 \to \infty$, where $\alpha _{\mathrm{sh}} \to 0$, also $\tilde{\alpha}{}_{\mathrm{sh}} \to 0$, and that for $R_0 \to 2 a r_0$, where the $v$-lines become light-like, $\tilde{\alpha}_{\mathrm{sh}} \to \pi$. By inserting (\ref{eq:shadowstat2}) into (\ref{eq:talpha2}) we find that
\begin{equation}
\mathrm{tan} ^2 \Big( \frac{\tilde{\alpha}{}_{\mathrm{sh}}}{2} \Big) = \dfrac{ R_0^3 (4 a r_0 - R_p) }{R_p^3 (R_0 - 2 a r_0)^2} \, \Bigg( 1 - \dfrac{(R_0-R_p)}{R_0} \sqrt{1 + \dfrac{2 a r_0 R_p}{R_0 \big( 4 a r_0 - R_p \big)}} \; \Bigg)^2 \, .
\end{equation}

\begin{figure*}[ht!]
\includegraphics[width = .45\textwidth]{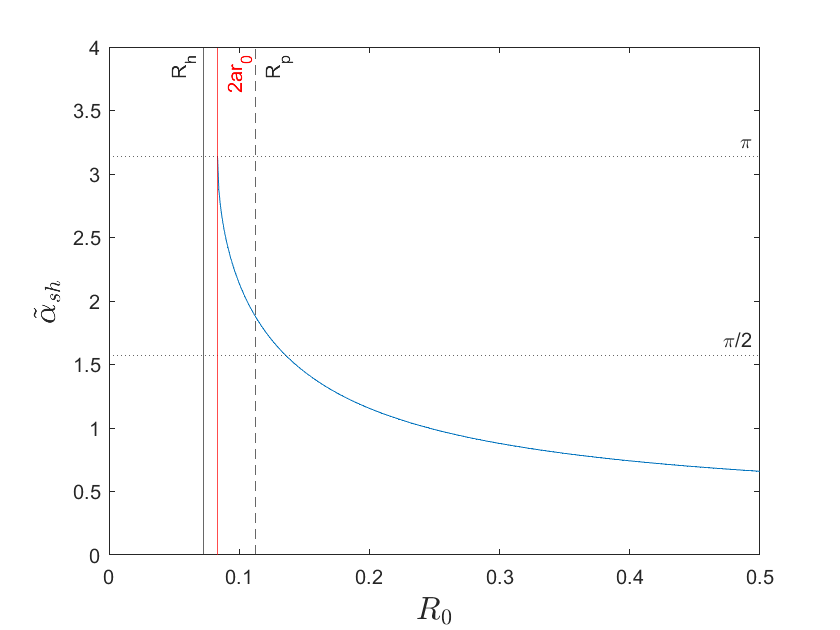} 
\caption{Angular radius of shadow as seen by observer on $ v $-line ($ a = 1/24 $)}
\label{shadow2}
\end{figure*}

The observer on a $v$-line starts at $R_0 = 2 a r_0$, asymptotically with the speed of light. The shadow, which in the beginning covers the entire sky, gradually shrinks while the observer moves outwards. It covers more than half of the sky at $R_0=R_p$, which reflects the fact that aberration has a magnifying effect in the backwards direction. The shadow vanishes for $R_0 \to \infty$.

\section{Conclusions}
\label{sec4}
In this paper we have considered the class of Vaidya metrics which describe the spherically symmetric and non-stationary space-time around a time-dependent mass. We have restricted ourselves to the special case that the mass function increases or decreases in such a way that the space-time admits a conformal Killing vector field. In this case the equation of light-like geodesics is completely integrable. In the case of an increasing mass function, the space-time gives us a model for an accreting black hole, whereas in the case of a decreasing mass function it can be considered as a (rough) model for a black hole that loses mass by way of Hawking radiation. In view of observations, the second case is certainly less relevant than the first because, for black holes of a Solar mass or more, the mass loss by Hawking radiation is very, very slow. However, we believe that both cases are of some interest from a conceptual point of view.

With the help of the conformal Killing vector field, we have explicitly calculated the light-like geodesics. In the case of a black hole with increasing mass, there are two horizons with a conformally static region in between, whereas in the case of a black hole with decreasing mass there is only one horizon, with the conformally static region between the horizon and infinity. We have found that in both cases there is a photon sphere in the conformally static region. In the case of increasing mass, the area of the horizons and of the photon sphere increases in the course of time, whereas in the case of decreasing mass the area of the horizon and of the photon sphere decreases in the course of time. The photon sphere determines the shadow. We have calculated the angular radius of the shadow for a conformally static observer (which has to be, of course, in the conformally static region), and we have found that it is time-independent, in spite of the fact that the area of the photon sphere is either increasing or decreasing with time. We have also calculated the red-shift under which one conformally static observer sees another one and we have found that it is time-independent. This is not in general true in conformally static metrics, as is exemplified by the well-known Robertson-Walker metrics,  but it is true here because the conformal factor has a special form. With the angular radius of the shadow known for conformally static observers, one can easily calculate the angular radius of the shadow for any other observer with the help of the aberration formula. We have exemplified this for some ingoing observers in the case of a black hole with increasing mass and for some outgoing observers in the case of a black hole with decreasing mass. 

What we have presented here is the third example of an exact analytical calculation of the shadow in a time-dependent situation. The other two examples are the shadow of an isolated collapsing star surrounded by vacuum \cite{2018} and  the shadow of a black hole in an expanding universe modeled by the Kottler metric \cite{PhysRevD.97.104062}.

\section*{ACKNOWLEDGMENTS}
VP gratefully acknowledges support from the Deutsche Forschungsgemeinschaft (DFG) within the Research Training Group 1620 ``Models of Gravity”.

\bibliographystyle{unsrt}

\bibliography{main}

\end{document}